\documentclass[aps,twocolumn,pra,reprint,amsmath,amssymb,floatfix,footinbib,superscriptaddress]{revtex4}

\usepackage{amsmath}	
\usepackage{gensymb}
\usepackage{hyperref}
\hypersetup{colorlinks=true}
\usepackage{upgreek}
\usepackage{graphics}
\usepackage{hyperref}
\usepackage{epsfig}
\usepackage{color}
\usepackage{bm}
\usepackage{float}
\usepackage{ulem} 
\usepackage{blindtext}
\usepackage{appendix} %

\usepackage{CJK}        
\usepackage{url}
\usepackage{multirow}

\newcommand{\upperRomannumeral}[1]{\uppercase\expandafter{\romannumeral#1}}

\usepackage{soul} 
\usepackage{color,xcolor}
\soulregister\cite7
\begin{document}

\title{Evidence of Kitaev interaction in the monolayer 1T-CrTe$_2$}

\author{Can Huang}
\affiliation{College of Physics, Nanjing University of Aeronautics and Astronautics, Nanjing, 211106, China}
\affiliation{Key Laboratory of Aerospace Information Materials and Physics, MIIT, Nanjing, 211106, China}
\author{Bingjie Liu}
\affiliation{College of Physics, Nanjing University of Aeronautics and Astronautics, Nanjing, 211106, China}
\affiliation{Key Laboratory of Aerospace Information Materials and Physics, MIIT, Nanjing, 211106, China}
\author{Lingzi Jiang}
\affiliation{College of Physics, Nanjing University of Aeronautics and Astronautics, Nanjing, 211106, China}
\affiliation{Key Laboratory of Aerospace Information Materials and Physics, MIIT, Nanjing, 211106, China}
\author{Yanfei Pan}
\affiliation{College of Physics, Nanjing University of Aeronautics and Astronautics, Nanjing, 211106, China}
\affiliation{Key Laboratory of Aerospace Information Materials and Physics, MIIT, Nanjing, 211106, China}
\author{Jiyu Fan}
\affiliation{College of Physics, Nanjing University of Aeronautics and Astronautics, Nanjing, 211106, China}
\affiliation{Key Laboratory of Aerospace Information Materials and Physics, MIIT, Nanjing, 211106, China}
\author{Daning Shi}
\email[]{shi@nuaa.edu.cn}
\affiliation{College of Physics, Nanjing University of Aeronautics and Astronautics, Nanjing, 211106, China}
\affiliation{Key Laboratory of Aerospace Information Materials and Physics, MIIT, Nanjing, 211106, China}
\author{Chunlan Ma}
\email[]{wlxmcl@mail.usts.edu.cn}
\affiliation{Jiangsu Key Laboratory of Micro and Nano Heat Fluid Flow Technology and Energy Application, School of Mathematics and Physics, Suzhou University of Science and Technology, Suzhou 215009, China}
\author{Qiang Luo}
\email[]{qiangluo@nuaa.edu.cn}
\affiliation{College of Physics, Nanjing University of Aeronautics and Astronautics, Nanjing, 211106, China}
\affiliation{Key Laboratory of Aerospace Information Materials and Physics, MIIT, Nanjing, 211106, China}
\author{Yan Zhu}
\email[]{yzhu@nuaa.edu.cn}
\affiliation{College of Physics, Nanjing University of Aeronautics and Astronautics, Nanjing, 211106, China}
\affiliation{Key Laboratory of Aerospace Information Materials and Physics, MIIT, Nanjing, 211106, China}

\date{\today}

\begin{abstract}
  The two-dimensional 1T-CrTe$_2$ has been an attractive room-temperature van der Waals magnet which has a potential application in spintronic devices. 
  Although it was recognized as a ferromagnetism in the past, the monolayer 1T-CrTe$_2$ was recently found to exhibit zigzag antiferromagnetism with the easy axis oriented at $70^\circ$ to the perpendicular direction of the plane.  
  Therefore, the origin of the intricate anisotropic magnetic behavior therein is well worthy of thorough exploration.
  Here, by applying density functional theory with spin spiral method, we demonstrate that the Kitaev interaction, together with the single-ion anisotropy and other off-diagonal exchanges, is amenable to explain the magnetic orientation in the metallic 1T-CrTe$_2$.
  Moreover, the Ruderman-Kittle-Kasuya-Yosida interaction can also be extracted from the dispersion calculations, which explains the metallic behavior of 1T-CrTe$_2$. 
  Our results demonstrate that 1T-CrTe$_2$ is potentially a rare metallic Kitaev material. 
\end{abstract}


\maketitle

\section{Introduction}
The Kitaev materials are drawing ever-growing attention for their potentially extraordinary magnetic properties, such as its capacity to offer quantum spin liquids \cite{Wen2019npj} or non-trivial spin textures, thereby serving as a hopeful material stage for the implementation of innovative applications in topological quantum computing and spintronics \cite{Rousochatzakis2023,Tokura2017,Trebst2021}. 
On a conceptual level, the seminal Kitaev honeycomb model \cite{Kitaev2006} was originally regarded as a toy model. The subsequent breakthrough was brought forward by Jackeli and Khaliullin \cite{Jackeli2009} through the interplay of the spin-orbit coupling (SOC) and electron correlation to materialize the Kitaev interaction. 
Sparked by their argument, extensive works have been concentrated on 
$j_{eff}$= 1/2 Mott insulators in the honeycomb lattices which are partially filled $4d$ and $5d$ shells of transition metal compounds, of which $\alpha$-RuCl$_3$ with zigzag (ZZ) ground state is the highly desirable Kitaev material \cite{Plumb2014,Banerjee2016,Ran2017}. 
The mixing of the crystal field and strong SOC results in the formation of a fully-filled $j_{eff}$ = 3/2 band and a half-filled $j_{eff}$= 1/2 band in the compounds.
Recent efforts have expanded the Kitaev system to two-dimensional high-spin configurations \cite{Lee2020,Cai2021prb}, notably in Cr-based monolayer hexagonal lattices \cite{Xu2018,Stavropoulos2019,Xu2020,Jaeschke-Ubiergo2021prb}. Furthermore, Kitaev materials are rare and rewarding in the exploration of metallic regimes.

Bulk van der Waals 1T-CrTe$_2$ is a ferromagnetic metal with a high Curie temperature of about 310K, which reaches the temperature required for practical spintronics applications successfully \cite{Freitas2015}. Synthesized epitaxial thin films of 1T-CrTe$_2$ in a metallic state also preserve a relatively high temperature down to the ultra-thin limit \cite{Zhang2021,Sun2021,Meng2021}.
Nevertheless, the latest atomic-resolution scanning tunneling microscopy experiments have proposed the antiferromagnetic (AFM) ZZ ground state of monolayer 1T-CrTe$_2$ with a 70-degree orientation to the perpendicular axis of the plane \cite{Xian2022}.
It has been demonstrated on real Kitaev materials that the competition between Heisenberg and Kitaev interaction stabilizes the ZZ structure, and furthermore the easy axis is related to the Kitaev anisotropy term \cite{Chaloupka2015,Sizyuk2014,Winter_2017}. 
Therefore, the origin of monolayer 1T-CrTe$_2$ with the easy axis oriented at $70^\circ$ in the vertical plane under the ZZ magnetic ordering is special and ambiguous, whereas the microscopic mechanism of this anisotropy in terms of bond-dependent interactions will be a superb study.

The 1T-CrTe$_2$ belongs to triangular-lattice structure \cite{Lv2015PRB}. 
The combination of geometric frustration and anisotropic exchange interactions induced by SOC has also proven to be a highly fruitful area.
In experiments, rare earth-based triangular lattice YbMgGaO$_4$ \cite{Li2015Sci,Li2015,Shen2016,Paddison2017} and others compounds \cite{Bordelon2019,Ding2019,Arh2022,Liu2018,Ortiz2023} have been intensively studied. 
Theoretically, triangular-lattice structures, accompanied with fractionalization and deconfinement, in certain regimes of their parameter space \cite{Li2015NewJ,Jackeli2015,Becker2015,Maksimov2019,Luo2017,Wang2021prb}. 
These studies point to the existence of Kitaev interaction within the triangular lattice, suggesting that they may also be present in 1T-CrTe$_2$. 
More importantly, hitherto, the Kitaev interaction that may be prevalent in systems containing heavy coordination elements, such as 1T transition metal triangular structures, have still not been discussed in the research process \cite{Xian2022,Li2021,Wu2022,Aghaee2022,Liu2022}.

In this work, we first point out, as far as we know, the existence of Kitaev physics in metallic 1T-CrTe$_2$ and elaborate the magnetization direction in the ZZ magnetic order off 70-degree from the perpendicular plane.
To begin with, using the spin spiral method, we perform DFT calculation to directly get the dispersion relations of the monolayer 1T-CrTe$_2$ based on generalized Bloch conditions.
In regard of the contribution of itinerant electrons to the long-range interactions within metals, we present the interactions among multiple nearest-neighbor (NN) atoms and apply it to the Heisenberg $J$, Kitaev $K$ and off-diagonal $\Gamma$ \cite{Chaloupka2015,Rau2014,Rousochatzakis2017}, except single-ion anisotropy (SIA) $A_\mathbf{k}$ term.
We quantitatively examine various magnetic Hamiltonian models and ultimately select the 3NN $J$-$K$-$\Gamma$-$A_{\textbf{k}}$ model.
The calculated spin spiral relations and the magnetic anisotropy energy (MAE) at different magnetic order are mapped into the optimal model to obtain the effective magnetic exchange parameters.
The angle of the magnetic easy axis mainly originated from the common competition between anisotropic Kitaev and SIA term.
Lastly, we find that the exchange parameter $\mathit{J_{i}}$ between different NN gained by the spin spiral relations follows a single Ruderman-Kittle-Kasuya-Yosida (RKKY) type interaction, indicating the role of the RKKY mechanism in the magnetic ordering.

The paper is organized as follows; we present computation details in Sec.~\ref{compu}.
 In Sec.~\ref{results}, we introduce the model of the monolayer 1T-CrTe$_2$ and analyze the anisotropic interactions in terms of the electronic and magnetic properties.
 We derive the NN spin model consisting of Heisenberg, Kitaev, and SIA interactions through the spin-spiral method and the MAE.
 Then we present our results on the calculated magnetic interactions and analyze the magnetization angle consistent with the Ref~\cite{Xian2022}. 
 We finally extract the RKKY interaction within this system.  
 We conclude in Sec.~\ref{conclusion}. 
 Details of  derivations in our work are given in the appendices.

\section{Computation Details}\label{compu}
We adopt the Vienna $ab\ initio$ simulation package (VASP), a software of first-principles pseudopotential plane wave method \cite{Kresse1999} based on Density Functional Theory (DFT), for calculation. 
The Perdew-Burke-Ernzerhof potential (PBE) \cite{Perdew1996} is used to form of the exchange correlation functional.
VASP can employ plane spiral basis functions to solve the Kohn-Sham equation through the self-consistent iterative method and calculate the force and tensor by the wave function.
1T-CrTe$_2$ is simulated by a slab model with periodic boundary conditions and along with a vacuum region of 20 \AA\ between adjacent slabs to avoid a spurious dipole moment from image supercells.
Atoms in the calculated system are relaxed to the ground state until the forces are less than $10^{-2}$ {eV/\AA}.
Energy cutoffs and k-points are chosen as 328 eV and $41\times41\times1$ k-point grids respectively for accurate calculations, which are much larger than that typically recommended.
The MAE of $\sqrt3\times2$ supercell is calculated statically using $19\times21\times1$ of k-points.
The convergence criterion for self-consistent total-energy calculations is ${1.0\times10^{-6}}$ eV.
\begin{figure}[!ht]
\centering
\includegraphics[width=0.95\columnwidth, clip]{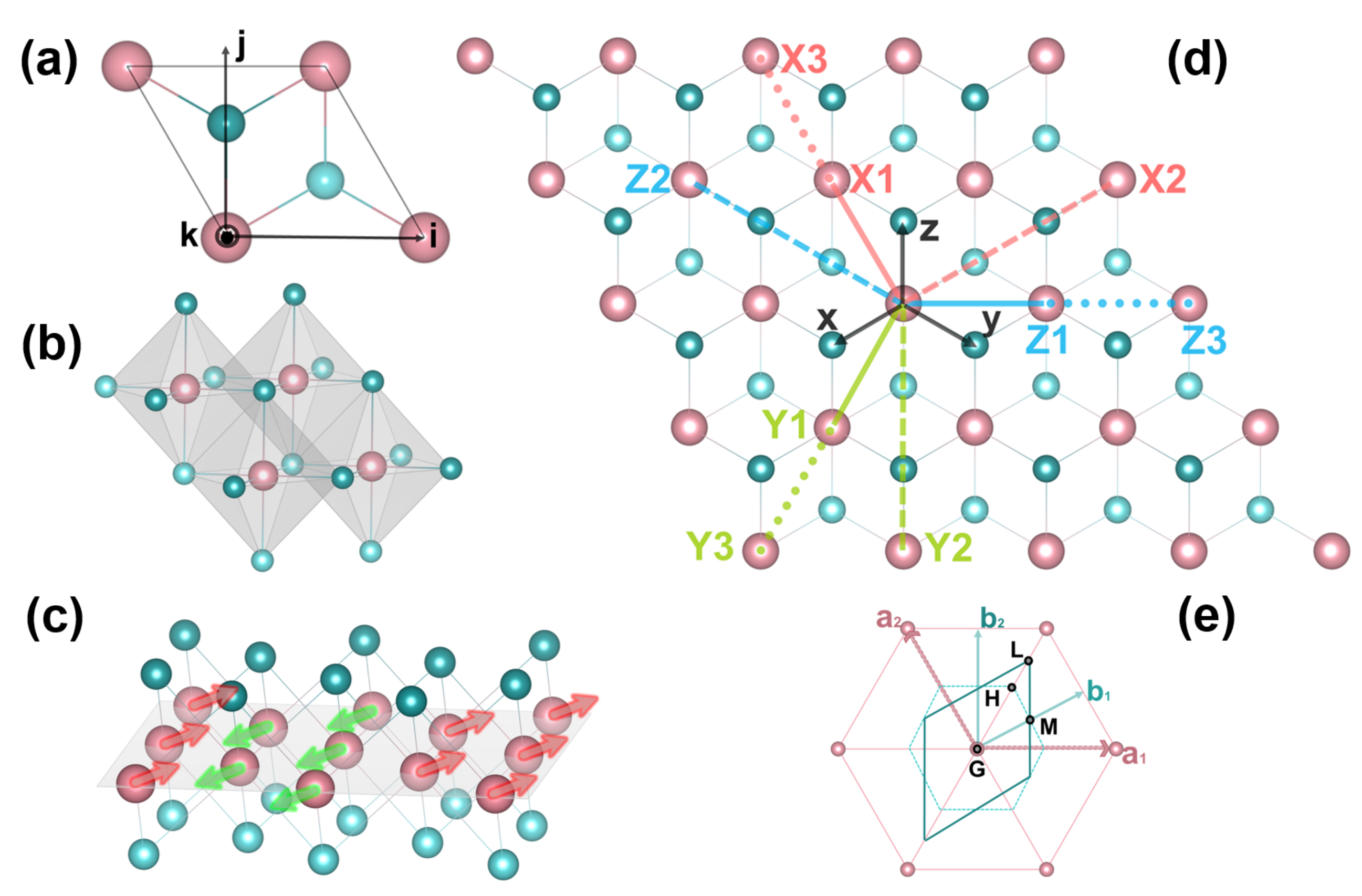}\\
\caption{(a) Top view of the monolayer 1T-CrTe$_2$. (b) The Octahedral coordination in 1T structure exhibites each Cr atom linked to six Te atoms. 
The dark (light) blue balls represent the top (bottom) Te atoms, and the pink balls represent the Cr atoms. (\textbf{i j k}) is the unit vector of the inter-perpendicular cartesian coordinate.
(c) Perspective view of the monolayer 1T-CrTe$_2$ under the ZZ magnetic ordering. Green and red arrows denote the magnetization directions of Cr atoms.
The light gray shaded surface indicates the $\textbf{ij}$ plane, and the orientation of the magnetic moment is at $70^\circ$ to the perpendicular plane direction, the {\textbf{k}}-axis.
(d) The ($X Y Z$) is the Cr-Cr bond, where the solid, dashed, and dotted lines label first-NN (1NN), second-NN (2NN), and third-NN (3NN) bonds on the triangular lattice, respectively. Blue, green, and red colors denote the $Z$, $Y$, and $X$ bonds, respectively.
(e) The Brillouin zone of the triangular lattice for 1T-CrTe$_2$, where $\mathbf{a}_1\ (1,\ 0,\ 0)$ and $\mathbf{a}_2\ ({-1/2},\ {\sqrt{3}/2},\ 0)$ are primitive lattice vectors. $\mathbf{b}_1$ and $\mathbf{b}_2$ are reciprocal lattice vectors. The ${\textbf{L}}$, ${\textbf{H}}$, ${\textbf{G}}$, and ${\textbf{M}}$ are the high symmetry points. 
}\label{FIG-Model}
\end{figure}
Methfessel-Paxton smearing with a half-width of 0.05 eV is used to accelerate the convergence for relaxed calculations and static calculations. 
In order to modify the strong Coulomb interaction associated with Cr 3$d$ orbit, the PBE+$U$ method with $U$ = 2.4 eV is applied \cite{Xian2022}. 
We also calculated the Heyd-Scuseria Ernzerhof (HSE06) hybrid functional \cite{Hummer2009} band structures for comparision with the PBE+$U$ band structures.
We perform fully noncollinear magnetic calculations within the projector-augmented wave formalism, as implemented in the VASP code by Hobbs et al. \cite{Hobbs2000} and SOC is also included in the present calculations. 
The optimized lattice constant of monolayer 1T-CrTe$_2$ is 3.698 \AA\
and in line with the recent theoretical results \cite{Xian2022,Liu2022}.
\section{Result}\label{results}
\subsection{ Model of 1T-CrTe$_2$}
The magnetic ions in 1T-CrTe$_2$ are located at the center of the edge-sharing octahedra, which is similar to the $\alpha$-RuCl$_3$ of conventional Kitaev materials. 
Moreover, recent DFT and many-body computation calculations point out that CrI$_3$ and CrSiTe$_3$ exhibit finite Kitaev interaction from the SOC of the heavy ligand elements Te/I \cite{Xu2018,Stavropoulos2019,Lee2020,Xu2020}. 
Ongoing efforts have extended Kitaev materials to Cr-based monolayer hexagonal lattices, and the Kitaev model has been applied to triangular lattices as well \cite{Kimchi2014,Becker2015,Catuneanu2015}. 
Furthermore, the latest experiments measured a stable ZZ oder of monolayer 1T-CrTe$_2$ and the magnetic easy axis is 70-degree off the {\textbf{k}}-axis \cite{Xian2022}. 
It therefore seems plausible to explore the Kitaev interaction in the monolayer 1T-CrTe$_2$.
More importantly, the monolayer 1T-CrTe$_2$ can keep ZZ magnetic state at a high temperature with strong magnetic anisotropy.
Mermin-Wagner theorem states that the presence of MAE is decisive for two-dimensional magnetic ordering. Xu et. al. noted that both SIA and Kitaev are responsible for determining the MAE \cite{Xu2018,Xu2020}. 
We shall revisit this anisotropy in monolayer 1T-CrTe$_2$ from the new perspective of bond-dependent interaction.
In particular, the microscopic mechanism of magnetic easy axis deviates at $70^\circ$ in the vertical plane is worth exploring. 

Fig.~\ref{FIG-Model}(a) and Fig.~\ref{FIG-Model}(b) show the top and side views of the monolayer with the Cr layer sandwiched between two Te layers. 
Each Cr atom is in the center of the octahedron formed by the nearest Te atoms, and Cr atoms are connected through the Cr-Te-Cr-Te plane. 
The ZZ magnetic configuration illustrated in Fig.~\ref{FIG-Model}(c) and the magnetic easy axis is $70^\circ$ off the {\textbf{k}}-axis. 
Afterwards, we verify that the magnetization angle can straightly capture the Kitaev interaction in Sec.~\ref{results}. 
According to the analysis of two Kitaev materials, CrI$_3$ and CrGeTe$_3$, the Kitaev model of each NN can be formed between the Cr atoms shown in Fig.~\ref{FIG-Model}(d) \cite{Xu2018}. 
The equatorial Cr-Te-Cr bond angles are $84.6^\circ$, which is close to $90^\circ$ and the Cr-Cr interaction should obey the Goodenough-Kanamori rules \cite{Kanamori1959,Goodenough1955}.

\subsection{ Basic Properties of 1T-CrTe$_2$ }
\begin{figure}[!ht]
\centering
\includegraphics[width=0.99\columnwidth, clip]{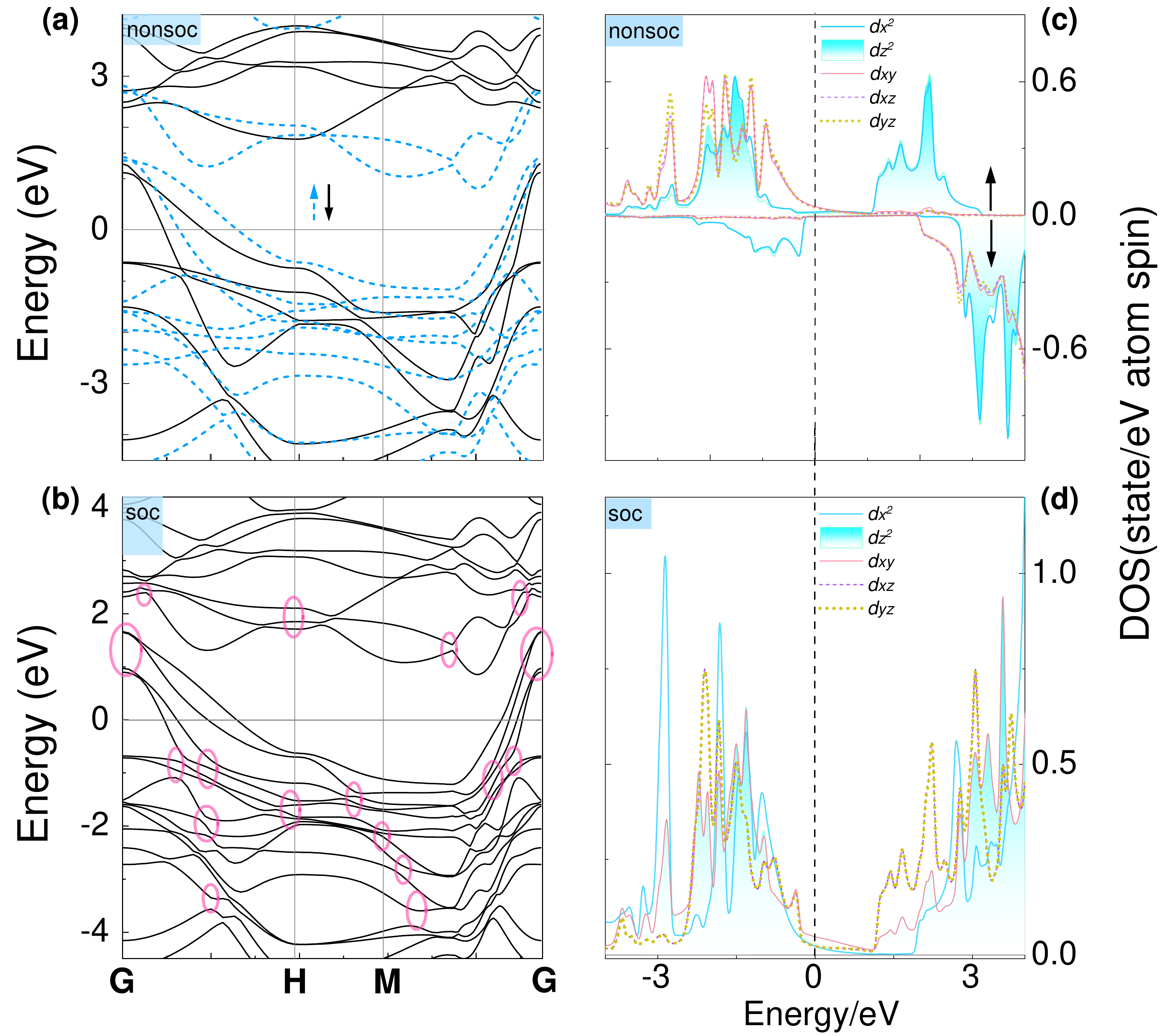}\\
\caption{(a)-(b) Band structure of monolayer 1T-CrTe$_2$ without and with SOC under FM order.
The blue and black bands are spin-up and spin-down channels in Fig.~\ref{FIG-band}(a), respectively. 
The pink circles represent the splitting of the formerly degenerate energy levels after the incorporation of SOC, as shown in Fig.~\ref{FIG-band}(b). 
(c)-(d) The density of states (DOS) of monolayer 1T-CrTe$_2$ without and with SOC.
  }\label{FIG-band}
\end{figure}

 The electronic structure of monolayer 1T-CrTe$_2$ in the absence and presence of SOC.
 Primarily, the spin-dependent electronic energy band structure without and with SOC under FM magnetic order configuration is investigated in Fig.~\ref{FIG-band}(a) and Fig.~\ref{FIG-band}(b), respectively.
 Both energy bands pass through the Fermi level and exhibit metallic properties.
In order to evaluate the band renormalization effect, we provide the HSE06 band structures in Fig. S1 of the supplementary material (SM) \cite{SuppMat}. 
The trend of the energy bands is roughly the same as that of the PBE+$U$ band, indicating that the band renormalisation is insignificant. Therefore, we mainly consider the PBE+$U$ method in the following calculation.
Notable spin splitting is observed when the SOC is turned on, particularly at the high symmetry sites. 
 These indicate that the magnetic interaction due to SOC can occupy a relatively important role, and this is the entry point to consider the Kitaev interaction within this system.
 The calculated $d$-orbital DOS projected to Cr are shown in Fig.~\ref{FIG-band}(c)-(d).
 The spin-up channel of the Cr atom is occupied by nearly all four $d$-electrons, while the states of the spin-down channel are almost empty. 
 Owing to the crystal field with the Cr atom located at the center of Te octahedral, the five degenerate $d$ orbitals split into threefold degenerated $t_{2g}$ and twofold degenerated $e_g$ orbitals as shown in Fig.~\ref{FIG-band}(c) and Fig.~\ref{FIG-band}(d).
 In the spin-up channel, the half-occupied states are $e_g$ orbitals from Cr-Te bonds. 
 The occupied parts are bonding orbitals, while the vacant parts occupied high energy zone are ani-bonding orbitals from the electrostatic repulsion. 
 In comparison, the fully occupied $t_{2g}$ orbitals lie between the Te ions and making them more relatively stable without the direct electrostatic repulsion between Cr and Te atoms.
 \subsection{ Model Hamiltonian}
 Actually, there exists not only Kitaev term in real materials but also other spin interactions, such as Heisenberg term and off-diagonal $\Gamma$ term. 
 In addition, for two-dimensional high-spin Kitaev materials and with heavy ligand elements, the SOC effect is manifested as the Kitaev-type interaction and SIA \cite{Xu2018,Stavropoulos2019,Xu2020,Jaeschke-Ubiergo2021prb,Riedl2022}.
Therefore, the minimal microscopic two-dimensional spin model of monolayer 1T-CrTe$_2$ has been chosen as the $J$-$K$-$\Gamma$-$A_{\textbf{k}}$. Namely, the Hamiltonian $\mathcal{H} = \mathcal{H}_{\mathrm{Hei}} + \mathcal{H}_{\mathrm{Ani}}$, where
\begin{align}\label{EQ:1Ham}
\mathcal{H}_{\textit{Hei}}=\sum_{i,j}\frac{1}{2}\left[J_{ij}\left(1-\mathbf{S}_i\cdot\mathbf{S}_j\right)\right]
\end{align}
and                 
\begin{align}\label{EQ:2Ham}
\mathcal{H}_{\textit{Ani}}&=\sum_{i,j}\frac{1}{2} \left[K_{ij}(S_{i}^{\gamma}S_{j}^{\gamma}-c^K_{ij})+\Gamma_{ij}(S_{i}^{\alpha}S_{j}^{\beta}-c^{\Gamma}_{ij})\right]  \notag \\
&\quad+\sum_i A_\mathbf{k}{\left[1-\big({(\mathbf{S}_i\cdot\mathbf{k})}^2\big)\right]}.
\end{align}
The term in Eq.~\eqref{EQ:1Ham} is the isotropic Heisenberg exchange interaction.
These terms in the Eq.~\eqref{EQ:2Ham} are in turn the Kitaev term, the off-diagonal interaction $\Gamma$, and the SIA term.
These coefficients characterize the constants of the interactions, where $c^K_{ij}$ and $c^{\Gamma}_{ij}$ are the numerical values of their respective interaction terms in the ferromagnetic states.
Here, $i$ and $j$ represent NN sites. The $\mathrm{\alpha}$, $\mathrm{\beta}$, and $\mathrm{\gamma}$ denote the spin direction on the $X$, $Y$, and $Z$ bond types, respectively.
$\mathbf{S}$ is the normalized spin operator, where its bond-dependent interaction has the spin component $\mathrm{\gamma}$ = $x$, $y$, $z$.
The triangular Cr layer lies in the $\textbf{ij}$ plane as shown in Fig.~\ref{FIG-Model}(d).
The SIA term is the last part and the $\mathbf{k}$-axis is the unit vector perpendicular to the plane, see Fig.~\ref{FIG-Model}(a). 
If ${A}_\mathbf{k}$ \textgreater 0, the easy magnetization axis is along the $\mathbf{k}$ direction. Otherwise, it lies in the $\textbf{ij}$ plane.

Next, we need to decompose the Kitaev term into various interaction parameters.
Especially, 1T-CrTe$_2$ is a metal system that has long-range RKKY interaction \cite{Ruderman1954,Kasuya1956}, while $\alpha$-RuCl$_3$ and CrSiGe$_3$ is a Mott-insulator \cite{Plumb2014} and semiconductor \cite{Williams2015}, respectively. 
The long-range interactions are more pronounced in two-dimensional magnetic materials than in the bulk, and their magnetic properties can generally be ascribed to RKKY interaction, which makes the description of magnetic states more complicating and challenging \cite{Balcerzak2007,Gong2017}.
\subsection{ Spin spiral method }
M. Marsman et al. \cite{Marsman2002} used an $ab\ initio$ spin-spiral approach for the total energy calculation with different magnetic structures of $\gamma$-Fe, which represented the initial work to obtain an effective magnetic exchange parameters $J_n$ with the framework of this numerical method. 
Understanding the salient features of magnetism in the monolayer 1T-CrTe$_2$, it is necessary to quantitatively analyze the magnetic interactions therein.
Precise calculation of multiple NN interactions should be needed, and it is confirmed in the spin spiral method based on generalized Bloch conditions \cite{Zhu2019,Zhu2020,Jiang2022,Huang2021}. 
All the individual models including RKKY, Kitaev, and SIA interaction can be disassembly by spin spiral method with generalized Bloch conditions. Hence, we can determine the coupling parameters of 1T-CrTe$_2$ in a certain range.
First, the magnetic Hamiltonian quantities associated with the present material are illustrated in Eq.~\eqref{EQ:1Ham} and Eq.~\eqref{EQ:2Ham}, and the individual equations of the magnetic Hamiltonian components can be derived via the spin-spiral methods combine with the generalized Bloch conditions. 
Second, the spin-spiral dispersion energy relations from a unit-cell are calculated, that is the total energy of the magnetic couplings.
Herein, DFT calculations with SOC for the noncollinear magnetic structure are the critical step to obtaining the respective parameters of Kitaev term.
Finally, by mapping the energy to each spin Hamiltonian equations, the parameters are obtained by using the least-squares method. 
Additionally, we calculate the spin-spiral dispersion energy of monolayer RuCl$_3$. 
The best-fit parameters for the $J_1$-$K$-$\Gamma$-$\Gamma'$ model give  $J_1$ = 1.68 meV, $K = -10.9$ meV, $\Gamma$ = 3.29 meV, and $\Gamma' = -0.28$ meV. 
As for the $J_1$-$J_3$-$K$-$\Gamma$ model, we find that $J_1$ = 0.36 meV, $J_3 = -0.27$ meV, $K = -11.0$ meV, and $\Gamma$ = 5.30 meV. 
Note that the sign of the Heisenberg interaction in Eq.~\eqref{EQ:1Ham} is opposite to the conventionally used symbol (e.g. see Ref~\cite{Rau2014}), which means that the third NN $J_3$ is physically antiferromagnetic.
In both models, our results suggest that the Kitaev and $\Gamma$ interactions are prominent and their signs are negative and positive, respectively. Those features are in accordance with the consensus achieved so far, showing the reliability of our method in determining the interaction parameters in Kitaev materials \cite{Winter_2017,Maksimov2020prr}.
The details of these calculations are placed in the Fig. S2 of the SM \cite{SuppMat}.

With generalized Bloch conditions, the magnetic moment $S(\mathbf{R}_j)$ at $j$ th-neighbor with wave vector $\textbf{q}$ is:
\begin{align}\label{EQ:3SPIN3SR}
{S}(\mathbf{R}_j)=S(0)\Big[\sin\left(\mathbf{q}\cdot\mathbf{R}_j+\frac{\pi}{4}\right)\mathbf{i}+\cos\left(\mathbf{q}\cdot\mathbf{R}_j+\frac{\pi}{4}\right)\mathbf{j}\Big].
\end{align}
The site of $j$ th-neighbor is depicted by $\mathbf{R}_j = m\mathbf{a}_1 + n\mathbf{a}_2 + l\mathbf{a}_3$, where $\mathbf{a}_1$, $\mathbf{a}_2$ and $\mathbf{a}_3$ are the basic vectors. 
The orientation of spin spiral is described as $\mathbf{q}=q_1\mathbf{b_1}+q_2\mathbf{b_2}+q_3\mathbf{b_3}$, of which $\mathbf{b}_1$, $\mathbf{b}_2$, and $\mathbf{b}_3$ are reciprocal lattice vectors as illustrated in Fig. \ref{FIG-Model}(e). 
All the magnetic moments are set in $\mathbf{ij}$ plane, with $S(0)$ forming an angle of 45° to $\mathbf{i}$-axis in our DFT calculations.
To avoid the impact of the Kitaev interaction by the Heisenberg term, we calculated the energy of $E(\mathbf{q})$ with and without SOC separately.
The detailed derivation equations are listed in Appendix~\ref{Appendix A}.
The Heisenberg interaction and Kitaev term with solely SOC can be separated by:
\begin{align}\label{EQ:4SN4SR}
E_S\left(\mathbf{q}\right)=E_{S+N}\left(\mathbf{q}\right)-E_N\left(\mathbf{q}\right)
\end{align}
and    
\begin{align}\label{EQ:5SN5SR}
E_S\left(\mathbf{q}\right)=E_{Hei_S}\left(\mathbf{q}\right)+E_{Kit}\left(\mathbf{q}\right).
\end{align}
The subscripts $S$ and $N$ denote the energy of the spin spiral relation with and without SOC, respectively.
$S+N$ represents the summation of both conditions. 
The subscripts $S$, $N$ and $S+N$ have the same implications for the following paragraphs, figures, and equations.
E$_{Hei_S}\left(\mathbf{q}\right)$ presents the energy of the Heisenberg interaction generated by the SOC only.
The summation of $E_{Hei_S}$ is as follows:
\begin{align}\label{EQ:6Heisenberg interaction_S}
E_{Hei_S}\left(\mathbf{q}\right)=E_{J_1}\left(\mathbf{q}\right)+E_{J_2}\left(\mathbf{q}\right)+E_{J_3}\left(\mathbf{q}\right).
\end{align}
Kitaev interaction is considered to the third NN, containing the $K$ and off-diagonal symmetric $\Gamma$.
The summation of $E_{Kit}$ is as follows:
\begin{align}\label{EQ:7Kitaevq}
E_{Kit}\left(\mathbf{q}\right)=\sum_{i=1}^{3}{{\big[E}_{K_i}\left(\mathbf{q}\right)}+E_{\mathrm{\Gamma}_i}\left(\mathbf{q}\right)\big] ,
\end{align}
where the subscripts $i$ represent 1NN, 2NN and 3NN. 
The summation of $E_S(\mathbf{q})$ is as follows:
\begin{align}\label{EQ:8Esq}
E_s\left(\mathbf{q}\right)=\sum_{i=1}^{3}{{\big[E}_{J_i}\left(\mathbf{q}\right)}+E_{K_i}\left(\mathbf{q}\right)+E_{\mathrm{\Gamma}_i}\left(\mathbf{q}\right)\big].
\end{align}
In case that SOC not considered, the system has only isotropic Heisenberg interaction. 
It is necessary to evaluate the $J_i$ for more distant nearest neighbors by considering the long-range magnetic sequence of the system:
\begin{align}\label{Eq:9NHeisenberg interaction}
 E_{Hei_N}\left(\mathbf{q}\right)=\sum_{i=1}^{15}{E_{J_i}\left(\mathbf{q}\right)} .
\end{align}
\subsection{ Magnetic anisotropy energy}
To explain the underlying magnetization angle, we use the unit-cell and $\sqrt3\times2$ supercell to calculate MAE, with the FM and ZZ AFM configurations, respectively.
The calculation of the MAE in \cite{Xian2022} was simulated and the angle was consistent, and the MAE of the FM ordering was also calculated. The angles $\theta$ and $\phi$ correspond to the angles between the magnetization direction and the $\mathbf{k}$ and $\mathbf{i}$ axes, respectively.
Under FM order, the magnetic moment $S(\mathbf{R}_j$) at $j$ th-neighbor with $\theta$ and $\phi$ is:
\begin{align}\label{EQ:10Ethetaq}
\mathbf{S}\left(\mathbf{R}_j\right)=S_0(\sin\theta \cos\phi\mathbf{i}+\sin\theta \cos\phi\mathbf{j}+\cos\theta\mathbf{k}).
\end{align}
Magnetic anisotropy is induced by SOC, where the angle of the magnetic easy axis is a combined contribution of the Kitaev and SIA terms.The detailed derivations are placed in Appendix~\ref{Appendix B}.
\begin{align}\label{EQ:11SIA}
E_{SIA}=A_\mathbf{k}\sin^2{\theta} ,
\end{align}
where $A_\mathbf{k}$ is the SIA term.  $E_{K_1}=E_{K_2}=E_{K_3}$ and $E_{\Gamma_1}=E_{\Gamma_2}=E_{\Gamma_3}=\frac{3}{2}\Gamma_i[-1+\cos{2\theta}]$.
The summation of $E^{FM}_{Kit}\big(\theta,\ \phi\big)$ is as follows:
\begin{align}\label{EQ:12FMk}
E^{FM}_{Kit}\big(\theta,\hspace{0.1cm}\phi\big)=\sum_{i=1}^{3}\big[E_{K_i}(\theta,\hspace{0.1cm}\phi)+E_{\Gamma_i}(\theta,\hspace{0.1cm}\phi)\big].
\end{align}
Among them, $K_1$, $K_2$, $K_3$, $\Gamma_1$, $\Gamma_2$, and $\Gamma_3$ are Kitaev parameters, including $S(0,\hspace{0.1cm}0)^2$.
Then, the total energy of MAE-FM $E^{FM}_{MAE}\big(\theta,\hspace{0.1cm}\phi\big)$:
\begin{align}\label{EQ:13FMk}
E^{FM}_{MAE}\big(\theta,\ \phi\big)=E_{SIA}+E^{FM}_{Kit}\big(\theta,\ \phi\big).
\end{align}
Under ZZ order, based on the Eq.~\eqref{EQ:10Ethetaq}, the initial magnetic moment is set at the  $\mathbf{i}\prime$ axis along the ZZ structure with $30^\circ$ angle to the $\mathbf{i}$ axis of the proto-cell structure.
All formulas are based on the coordinate frame ($\mathbf{i\hspace{0.1cm}j\hspace{0.1cm}k}$) of the proto-cell.
The summation of $E^{ZZ}_{Kit}\big(\theta,\ \phi\big)$ is as follows:
\begin{align}\label{EQ:14ZZk}
E^{ZZ}_{Kit}\big(\theta,\ \phi\big)=\sum_{i=1}^{3}\big[E_{K_i}(\theta,\ \phi\big)+E_{\Gamma_i}(\theta,\ \phi)\big] 
\end{align}
Among them, $K_i$ and $\Gamma_i$ represent parameters of different NN respectively, including ${S(0,\ \pi/6)}^2$.
Then, the total energy of ZZ-MAE $E^{ZZ}_{Kit}\big(\theta,\ \phi\big)$ is: 
\begin{align}\label{EQ:15ZZktotal}
E^{ZZ}_{Kit}\big(\theta,\ \phi\big)=E_{SIA}+E^{ZZ}_{Kit}\big(\theta,\ \phi\big),
\end{align}
\subsection{ Kitaev interaction parameters}
\begin{figure*}[!ht]
\centering
\includegraphics[width=2\columnwidth, clip]{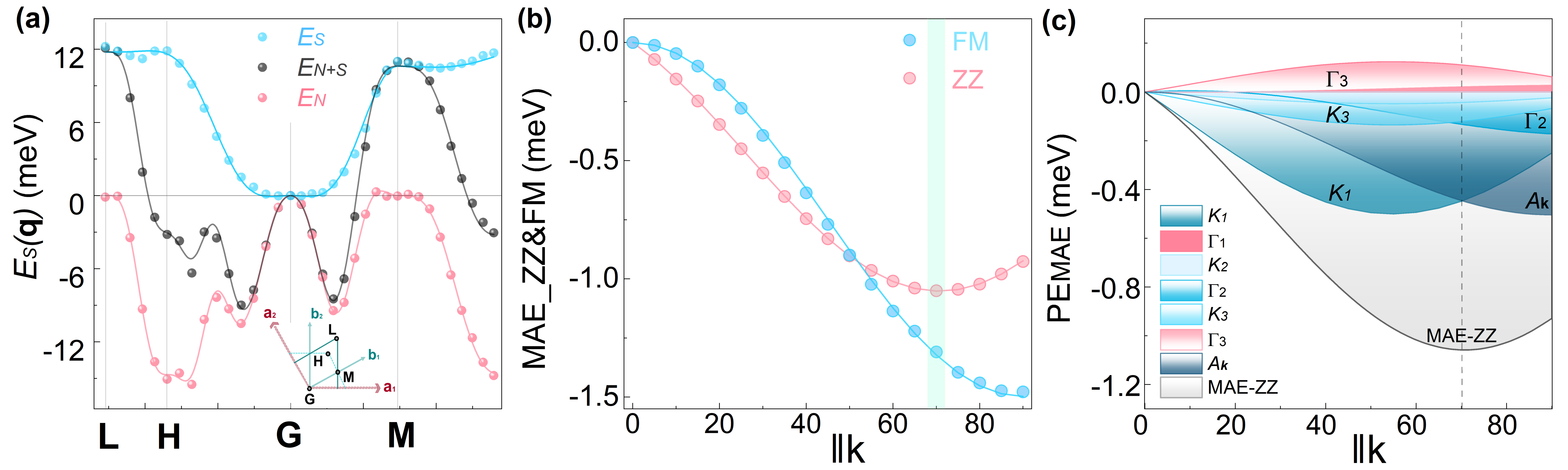}\\
\caption{(a) $E(\mathbf{q})$ as a function of the spiral wave vector $\mathbf{q}$ in the monolayer 1T-CrTe$_2$.
The “$E_S$” denotes the energy with SOC and the “$E_N$” is the energy without SOC, which represented by blue and pink spheres, respectively.
The total energy of “$E_{N+S}$” is emphasized by blue spheres. $\mathbf{L}$, $\mathbf{H}$, $\mathbf{G}$, and $\mathbf{M}$ are special k-points in the Brillouin zone shown in the inset graph.
(b) The MAE of the unit-cell with FM order and the $\sqrt3\times2$ supercell with ZZ order, which are labeled with pale blue and pale pink circles, respectively.
In Fig.~\ref{FIG-eqall}(a) and Fig.~\ref{FIG-eqall}(b), scaled symbols are calculated results while lines are fitted ones.
(c) Partial energy of the MAE (PE$_{MAE}$) of ZZ order with the magnetic angle. 
The parameters that make PE$_{MAE}$ greater than 0 meV contain $\Gamma_1$ and $\Gamma_3$, as shown in the rose-pink color family.
The parameters that make PE$_{MAE}$ less than 0 meV contain $K_1$, $K_2$, $\Gamma_2$, $K_3$, and $A_\mathbf{k}$, as shown in the lapislazuli color family.
The MAE of the ZZ magnetic order is represented by the silver color.
  }\label{FIG-eqall}
\end{figure*}

The SOC is an essential factor that causes the Kitaev and SIA interaction. The Heisenberg, Kitaev, and SIA term with SOC are mixed in the energy of $E_S$(\textbf{q}).
In Fig.~\ref{FIG-eqall}(a), $E_{N+S}$(\textbf{q}) and $E_N$(\textbf{q}) are calculated results, and $E_S$(\textbf{q}) is gotten based on Eq.~\eqref{EQ:3SPIN3SR}. 
The lowest energy point of $E_{N+S}$(\textbf{q}) is a spin spiral AFM state.
The ZZ state does not belong to $\mathbf{q}$ which is described by spin spiral relations. 
Then, we compute the energy under FM and ZZ magnetic order individually by using a unit-cell and $\sqrt3\times2$ supercell.
The energy difference between these two structures is $-17.1$ meV/Cr, which is much smaller than the minimum energy of $E_S$(\textbf{q}). 
Hence, the ZZ structure is still the ground state in our calculations.
The schematic arrangement of the magnetic moments at the high symmetry and the lowest energy points in Fig.~\ref{FIG-eqall}(a) are put in the Fig. S3 of the SM \cite{SuppMat}. 
Here is a very important $E_{N+S}$(\textbf{q}) feature that reflects the existence of Kitaev interaction, which suggests that it can be further verified with neutron scattering experiments \cite{Zhang2018}.
As shown in Fig.~\ref{FIG-Model}(e) and Fig.~\ref{FIG-eqall}(a), due to $C_6$ rotational symmetry, $E_N$(\textbf{L}) and $E_N$(\textbf{M}) are equal based on Eq.~\eqref{EQ:3SPIN3SR} and has isotropic Heisenberg exchange interaction alone. 
The $C_6$ symmetry bareaking under the Kitaev interaction leads to $E_S$(\textbf{L}) not being equal to $E_S$(\textbf{M}), according to Eqs. \eqref{EQ:6Heisenberg interaction_S}-\eqref{EQ:8Esq}.
Of course, $E_{N+S}$(\textbf{L}) is not equal to $E_{N+S}$(\textbf{M}) either, and their difference can be characterized in the experiment as the strength of the Kitaev interaction.

For the case without SOC, the system has only isotropic Heisenberg interactions, for which the detailed parameters $J_{Ni}$ are given in Table~\ref{Table1}. 
We select the coupling parameters up to the $15th$ nearest neighbor to represent the long-range interaction. 
The detailed parameters with SOC in Table~\ref{Table2} are obtained by fitting the calculated values in Fig.~\ref{FIG-eqall}(a)-(b) based on Eqs.~\eqref{EQ:6Heisenberg interaction_S}-\eqref{EQ:8Esq} and Eqs.~\eqref{EQ:10Ethetaq}-\eqref{EQ:15ZZktotal}. 
In this case, Eqs.~\eqref{EQ:6Heisenberg interaction_S}-\eqref{EQ:8Esq} represent the Kitaev interaction and off-diagonal symmetric $\Gamma$ extracted from the dispersion relation $E_S(\mathbf{q})$ in Fig.~\ref{FIG-eqall}(a).
Meanwhile, we calculated the MAE of FM and ZZ magnetic configurations respectively in Fig.~\ref{FIG-eqall}(b), in which the fitted lines are obtained based on Eqs.~\eqref{EQ:10Ethetaq}-\eqref{EQ:15ZZktotal}.
\begin{table*}[htbp]
  \centering
  \caption{In the absence of SOC, the obtained isotropic Heisenberg interaction interaction parameters of monolayer 1T-CrTe$_2$. All parameters $J_{iN}$ were obtained by least-squares fitting based on the calculated spiral dispersion energy $E_N(\mathbf{q})$ and Eq.~\eqref{Eq:9NHeisenberg interaction}.} \label{Table1}
  \begin{ruledtabular}
 \label{Table1}
    \begin{tabular}{ c|c c c c c c c c c c c c c c c}
      meV & $J_1$ & $J_2$ & $J_3$ & $J_4$ & $J_5$ & $J_6$ & $J_7$ & $J_8$ & $J_9$ & $J_{10}$ & $J_{11}$ & $J_{12}$ & $J_{13}$ & $J_{14}$ & $J_{15}$ \\
      \hline
      $N$ & -0.228 & 1.635 & -2.864 & -0.443 & -0.477 & 0.111 & -0.100 & 0.539 & 0.210 & -0.076 & -0.070  & -0.098 & -0.018 & 0.030 & 0.056 \\
    \end{tabular}
    \end{ruledtabular}
\end{table*}
\begin{table}[htbp]
  \centering
  \caption{In the case of containing SOC, the isotropic Heisenberg interaction and anisotropic interaction parameters are obtained for monolayer 1T-CrTe$_2$. 
  The $E_S(\mathbf{q})$ was separated by Eqs.~\eqref{EQ:4SN4SR}-\eqref{EQ:5SN5SR} to ensure that the system contains only the Heisenberg interaction parameters $J_S$ caused by SOC and the anisotropic $K$ term of Kitaev, $\Gamma$ term, together with the $A_\mathbf{k}$ term of SIA. The values of $J_{1S}$, $J_{2S}$, and $J_{3S}$ are 2.685 meV, $-0.384$ meV, and $-0.607$ meV, respectively. } \label{Table2}
   \begin{ruledtabular}
 \label{Table2}
    \begin{tabular}{ c| c c c c c c c c }
      meV & $K_1$ & $K_2$ & $K_3$ & $\Gamma_1$ & $\Gamma_2$  & $\Gamma_3$  & $A_\mathbf{k}$ & \\
      \hline
      $S$  & -1.501 & -0.021 & 0.145 & 0.129 & 0.203 & 0.186 & -0.505 &  \\
    \end{tabular}
     \end{ruledtabular}
\end{table}
Fitted lines are matched very well to the calculated results in Fig.~\ref{FIG-eqall}(a) and Fig.~\ref{FIG-eqall}(b), which shows the validness of these obtained exchange parameters.
$J_{3N}$ has the largest value of $-2.864$ meV in Table~\ref{Table1}, and $J_{1N}$ is an order of magnitude smaller than $J_{3N}$.
In parallel, the long-range magnetic interactions of this system can exhibit RKKY interaction due to the metallic character of the monolayer 1T-CrTe$_2$ in Fig.~\ref{FIG-band}. 
Notice that $J_{9N}$ makes a contribution to keeping the monolayer magnetic ordering up to a value of 0.210 meV.
The value of $J_{N>9}$ is very tiny, which is two orders of magnitude smaller than $J_{3N}$. 
After adding SOC, $J_{1S}$ results in an equivalent value to that of $J_{3N}$. It is an order of magnitude larger than $J_{2S}$ and $J_{3S}$. 
According to Eq.~\eqref{EQ:4SN4SR}, the Heisenberg interaction in $E_{N+S}(\mathbf{q})$ is the sum of the contributions with and without SOC, denoted by $J_{N+S}$. 
$J_{N+S}$ is the sum of $J_N$ and $J_S$, which represents the overall Heisenberg interaction. 
\begin{figure*}[!ht]
\centering
\includegraphics[width=2\columnwidth, clip]{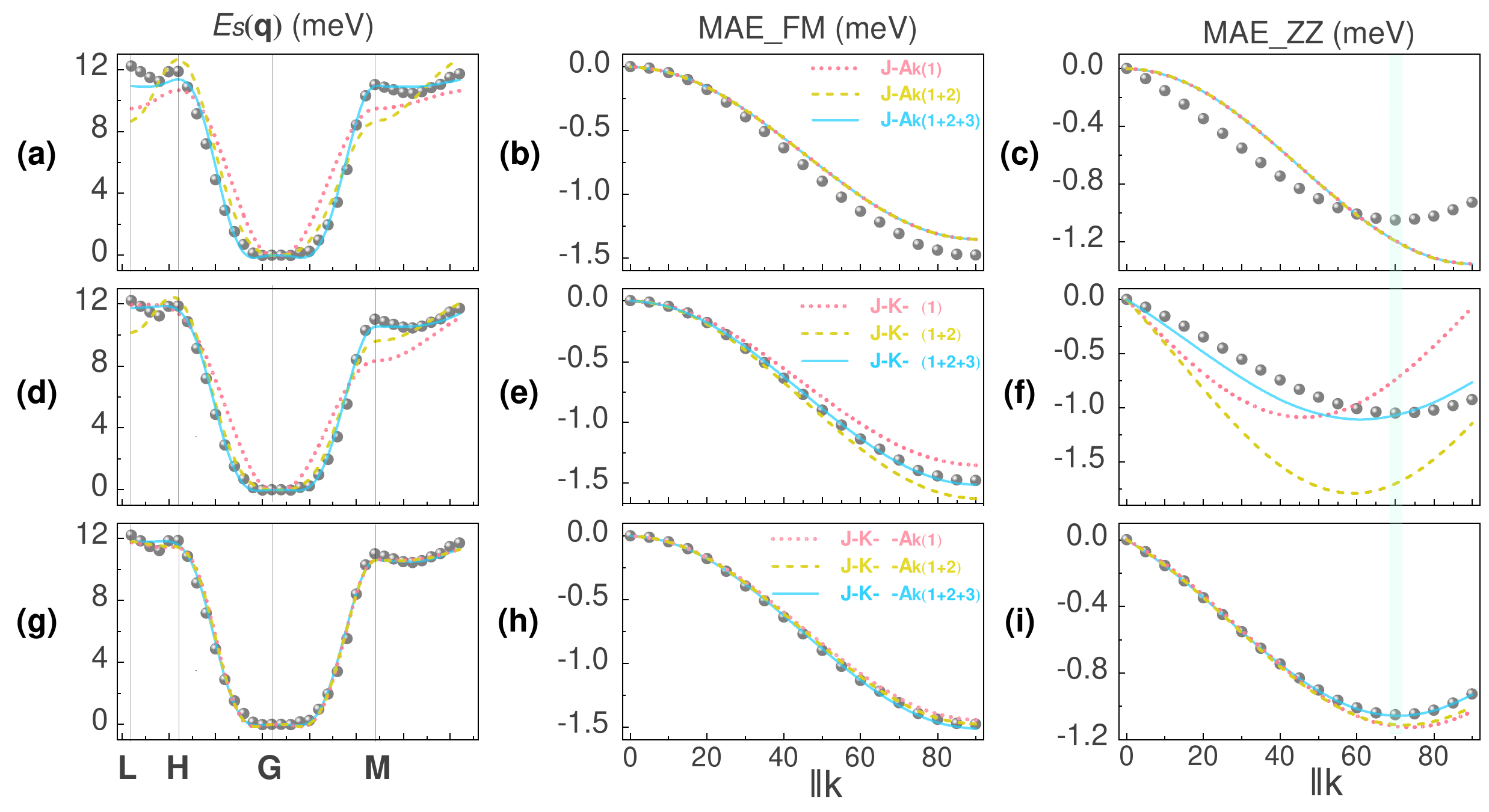}\\
\caption{Each column corresponds to the $E(\textbf{q}$), MAE-FM, and MAE-FM fitted panels, respectively, and the grey spheres represent the computed values. 
The pink, mustard, and blue lines represent models fitting the performance diagram by considering the 1NN, considering to the 2NN, and considering to the 3NN, respectively. 
(a)-(c) The first rows present, in order, the $J$-$A_\textbf{k}$ model using different neighbors, that is, with only the Heisenberg and SIA term in the model.
(d)-(f) stand for the $J$-$K$-$\Gamma$ model. 
(g)-(i) Represent the $J$-$K$-$\Gamma$-$A_\textbf{k}$ model.
  }\label{FIG-nhmodel}
\end{figure*}

In the following, we concentrate on Kitaev physics.
The Kitaev interaction attains a value of $-1.501$ meV, which is comparable in magnitude to $J_{1S}$.
The magnitude of $K_1$ is approximately twice times larger compared to that of the monolayer CrI$_3$ \cite{Xu2020prb} and four times larger than the monolayer CrSiTe$_3$ \cite{Xu2020}. 
Moreover, the Heisenberg interaction of the monolayer 1T-CrTe$_2$ is more robust, which is consistent with previous studies on Cr-based Kitaev materials \cite{Xu2020,Stavropoulos2021prr}.
In addition, this peculiar magnetization angle offers a possibility for the Kitaev interaction. 
The ZZ structure presents an angle of $70^\circ$ between the magnetic moment to the $\mathbf{k}$-axis \cite{Xian2022}, and our calculations are consistent with it, as shown in Fig.~\ref{FIG-eqall}(b) and Fig.~\ref{FIG-eqall}(c). 
This angle is closely related to the arrangement of the magnetic moments of the ZZ structure. 
If the magnetic moments are arranged ferromagnetically, the magnetization axis is in the $\mathbf{ij}$ plane [see Fig.~\ref{FIG-eqall}(b)].
These findings further confirm the presence of the Kitaev interaction in 1T-CrTe$_2$.

To examine deeply the effect of Kitaev and SIA factors in determining the magnetization angle under the ZZ configuration, we present a PE$_{MAE}$ diagram in Fig.~\ref{FIG-eqall}(c), which illustrates the variation of MAE with angle for each parameter. 
Notably, the magnetisation axis tends to align out-of-plane with the $\textbf{k}$ direction primarily owing to the cooperation of $K_1$, $K_3$, and $\Gamma_2$.
In this case, $A_\textbf{k}$ and $\Gamma_2$ make the ZZ magnetic moments favor in-plane, while $K_1$ and $K_3$ make the ZZ moments point to $55^\circ$ away from the $\textbf{k}$-axis. 
As a reslut, if only SIA exists in 1T-CrTe$_2$, the magnetic moment aligns in parallel or perpendicular to the $\mathbf{k}$-axis. To sum up, a combination of the anisotropic Kitaev interaction and SIA results in the magnetic moment with a 70-degree off the $\mathbf{k}$-axis.

Compared with aforementioned monolayer Kitaev materials \cite{Xu2020prb,Xu2020}, 1T-CrTe$_2$ demonstrates a relatively stronger Kitaev interaction and becomes a competitive Kitaev material.
The next question is how to suppress the Heisenberg interaction in 1T-CrTe$_2$ in order to achieve Kitaev spin liquid state. Indeed, we aim to induce a potential Kitaev spin liquid by applying strain to suppress non-Kitaev interactions. 
However, it was found that the Kitaev and other interactions decreased simultaneously when strain was applied to the compression of the 1T-CrTe$_2$, and vice versa. 
Accordingly, unlike CrSiTe$_3$ \cite{Xu2020}, the achievement of Kitaev spin liquid in the monolayer 1T-CrTe$_2$ by applying strain solely is challenging and worthy of subsequent explorations.

\begin{figure*}[!ht]
\centering
\includegraphics[width=2\columnwidth, clip]{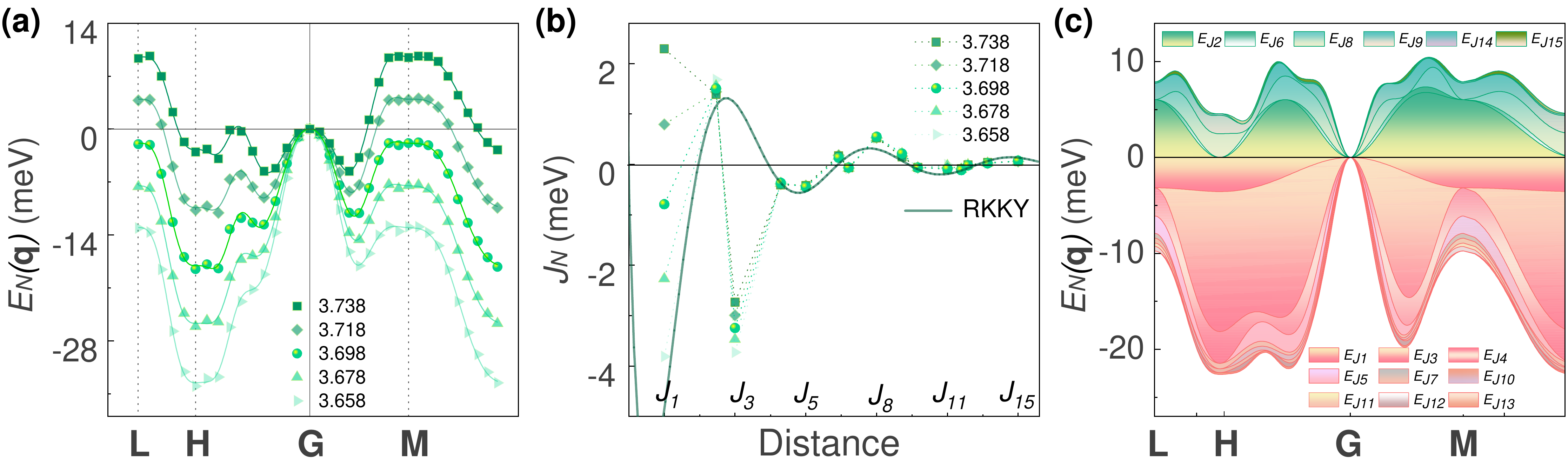}\\
\caption{(a) scattered symbols are $E_N(\textbf{q})$ for varying lattice constants $a_0$ (shown in the legend with length in \AA); the lines are fitted with the Heisenberg interaction $J$ parameter in (b).
(b) Heisenberg interaction $J_N$ for each lattice constant is fitted with the least-square methods; the dark green line is the fitting of selected Heisenberg interaction $J$ $(J_{5,6,8\dots15}$) with RKKY model.
(c) The stacked area chart of $E_N(\textbf{q})$ for $J_N$ as $a=3.698$ \AA.
  }\label{FIG-arkky}
\end{figure*}

In this work, we choose the effective model to fit all the calculated dispersion relations of $E(\textbf{q})$ and the MAE under different magnetic ordering simultaneously.
If only the Heisenberg and SIA terms are regarded as the relevent model to describing magnetic interactions, the simplest MAE-FM cannot be fit despite accounting for the third-NN[see Fig.~\ref{FIG-nhmodel}(a)-(c)]. 
To assess the significance of the SIA term, we employ the $J$-$K$-$\Gamma$ model.
As depicted in Fig.~\ref{FIG-nhmodel}(d)-(f), when third-NN is considered, the fitted line can already match $E_S$(\textbf{q}) and MAE-FM, yet it still fails to match MAE-ZZ, and there is still a considerable discrepancy.
Ultimately, these calculated curves can be fit rather successfully only when considering all terms altogether and fitting to the third-NN, as shown in Fig.~\ref{FIG-nhmodel}(g)-(h).
Without term $J$, only the Kitaev and SIA terms are considered, as in SM \cite{SuppMat} Fig. S4, which also does not allow fitting all data simultaneously.
Most remarkably, in SM ~\cite{SuppMat} Fig. S5, when fitting MAE-ZZ singularly, the fitted parameters will seriously deviate from the calculated values of MAE-FM. 
With this in mind, it is important to identify a model that is appropriate for monolayer 1T-CrTe$_2$ and requires consideration of longer-range interactions. Simultaneously, we ought to comprehensively evaluate numerical and experimental results while avoiding the obtaining of magnetic parameters from a single numerical or experimental result.
\subsection{RKKY interaction}
As shown in  Table~\ref{Table1}, we have already gotten the long-range magnetic interactions because of the metallicity of 1T-CrTe$_2$, and these $J_N$ can be described by the RKKY mechanism in our previous work \cite{Zhu2019}. 
RKKY model can be mapped onto the classical Heisenberg model \cite{Prange1979}. $J_{iN}$ can be used in the RKKY model directly and in two-dimensional structure, this model can be expressed as \cite{Zhu2023}:
\begin{align}\label{EQ:15rkky}
J_{iN}\left(r\right)=a\frac{\cos({7.62}\sqrt n d)}{d^2} .
\end{align}
The $a$ is a constant, $d=r/a_0$ (where $r$ is the distance between magnetic atoms and $a_0$ is the lattice constant of 1T-CrTe$_2$) is the scaled distance, and $n$ is the electronic number in a unit cell to mediates the RKKY interaction.
If $n$ is a constant for $a_0$, then the $J_{iNs}$ of RKKY are irrelevant to $a_0$.
Eq.~\eqref{EQ:15rkky} gives us a way to pick up the $J_{iNs}$ belong to RKKY by DFT calculations. 
Various lattice constants $a_0$ can be chosen and $J_{iNs}$ obtained at each lattice constant, as shown in Fig.~\ref{FIG-arkky}(a) and (b). 
If these calculated $J_{iNs}$ are not varying with the lattice constant $a_0$, they fall under the RKKY interaction. 
As depicted in Fig.~\ref{FIG-arkky}(b), these $J_{iNs}$ show an oscillatory behavior. 
These $J_{1N}$ shoot up strongly as the lattice constant $a_0$ increases, while the ascent of $J_{3N}$ are relatively modest. 
In Fig.~\ref{FIG-arkky}(b), we can also note that the other $J_{iNs}$ hardly change with $a_0$. Then, they are substituted into Eq.~\eqref{EQ:15rkky} for fitting the RKKY interaction. 
For the monolayer 1T-CrTe$_2$, the RKKY interaction is mediated by 0.17 electrons per unit cell.
Under the trend of RKKY interaction, the NN Heisenberg interaction exhibits the nature of AFM.
And its absolute value is larger than the actual calculated $J_{1N}$ as seen in Fig.~\ref{FIG-arkky}(b). 
Moreover, the differences between $J_{1N}$ and RKKY are all positive. This makes the system likely to form a FM state and also satisfies the Goodenough-Kanamori rule \cite{Kanamori1959,Goodenough1955}. 
The contribution of each nearest neighbor $J_Ni$ to $E_N$(\textbf{q}) , as shown in Fig.~\ref{FIG-arkky}(c).
Negative $J_{1N}$ and $J_{3N}$ render the AFM system, while positive $J_{2N}$ is ferromagnetic. 
By increasing $a_0$, not only $J_{1N}$ grows strongly but also the ground state of this system turns form AFM to FM, as shown in Fig.~\ref{FIG-arkky}(a) and (b).  
Consequently, this provides an explanation for the previous work which measured that the monolayer 1T-CrTe$_2$ is ferromagnetic \cite{Zhang2021}, precisely due to the sensitivity of the lattice constant to the magnetic properties of the system.
\section{Conclusion}\label{conclusion}
To summarize, we present a comprehensive analysis of the intrinsic Kitaev interaction, electronic and magnetic properties of the metallic monolayer 1T-CrTe$_2$ by DFT calculations.
We employ the spin spiral method with generalized Bloch conditions to extract the spin spiral dispersion relation $E(\textbf{q})$ for the monolayer 1T-CrTe$_2$. 
In Fig.~\ref{FIG-eqall}(a), the $E(\textbf{q})$ shows the breaking of $C_6$ symmetry induced by the SOC and indicates the existence of anisotropic Kitaev interaction. 
Meanwhile, the MAE is different between FM and ZZ order, and this phenomenon cannot be described properly by SIA alone.
We also revisit the magnetic anisotropy of monolayer 1T-CrTe$_2$ and elucidate the magnetization angle off the $\mathbf{k}$-axis by 70-degree at the ZZ order. 
We carefully choose the optimal model containing the Heisenberg $J$, Kitaev coupling $K$, symmetric off-diagonal exchange $\Gamma$ and SIA $A_\textbf{k}$ term to determine magnetism in 1T-CrTe$_2$.
Finally, we pin down the magnetic interaction parameters by using the calculated $E(\textbf{q})$ and the MAE at different magnetic order settings.
In Fig.~\ref{FIG-nhmodel}, all these terms are required to allow for third-NN except SIA, which reveals the contribution of itinerant electrons to the long-range interactions in metals.

In detail, we discover that a dominant Heisenberg interaction $J_1$ and a finite Kitaev interaction can maintain the magnetic ordering in the monolayer limit. 
An important finding is the Kitaev term, especially the $K_1$ term has a value of $-1.501$ meV, which is approximately more than 4 times larger than CrSiTe$_3$ \cite{Xu2020}.
And the competition of $K_1$ and SIA term jointly leads to a magnetization angle of $70^\circ$ on the $\textbf{k}$-axis. 
From DFT total energy calculations, we demonstrate that the ZZ magnetic order of 1T-CrTe$_2$ is more stable. 
Notably, the RKKY mechanism of 1T-CrTe$_2$ can also be characterized by the interaction parameters for its metallic long-range magnetic interaction.
As shown in Fig.~\ref{FIG-arkky}, the obtained Heisenberg $J_i$ do not vary with the lattice constants, which is considered to belong to the RKKY interaction and has oscillating behavior. 
Moreover, the lowest energy of monolayer 1T-CrTe$_2$ rises as the lattice constant increases, which gives light on the way for subsequent work to explore the influence of the lattice constant on the ground state and magnetic interaction in the monolayer limit of the system.
Besides, we infer that there are two essential elements for finding bond-dependent Kitaev interaction in two-dimensional magnetic materials.
One is the transition metal cation located at the center of a shared octahedron around a triangular lattice, while the other is the anion that is a heavy ligand. 

Our findings explain the mechanism of magnetic interactions in the monolayer 1T-CrTe$_2$ from a microscopic perspective. 
Although the Kitaev interaction is finite and achievement of Kitaev spin liquid in monolayer 1T-CrTe$_2$ is challenging however still worthy of a follow-up explorations.
With this insight, our work provides a theoretical basis for further research on the bond-dependent interactions in two-dimensional van der Waals materials. 
\begin{acknowledgements}
This work was supported by the National Natural Science Foundation of China (NSFC) with Grants No. 11204131 and No. 12247183.This work is partially supported by High Performance Computing Platform of Nanjing University of Aeronautics and Astronautics.
\end{acknowledgements}
\begin{widetext}
\appendix
\setcounter{equation}{0} 

\section{Spin spiral relationship of individual paramerters} \label{Appendix A}
After adding SOC, not only Heisenberg interaction generated by SOC but also other anisotropic interactions should be regarded based on Eq.~\eqref{EQ:4SN4SR}. 
We set the magnetic moment in the plane after the inclusion of SOC, for which the anisotropic interaction contains solely the Kitaev interaction.
Kitaev interaction is considered to the third NN of Eq.~\eqref{EQ:5SN5SR}, containing the off-diagonal symmetric $\Gamma$.
The summation of Kitaev between $S(0)$ and $S(\mathbf{R}_j)$ is:
\begin{align}
 E_{K_1}(\mathbf{q})&=\frac{1}{2}K_1\bigg[{\cos}^2 {\left(\frac{7\pi}{12}\right)}\cos2\pi \left(q_1+q_2\right)+{\cos}^2 {\left(\frac{\pi}{12}\right)}\cos2\pi q_2+{\cos}^2 {\left(\frac{3\pi}{4}\right)}\cos2\pi q_1-3\bigg] , \notag  \\
E_{K_2}(\mathbf{q})&=\frac{1}{2}K_2\bigg[{\cos}^2 {\left(\frac{7\pi}{12}\right)}\cos2\pi \left(q_1-q_2\right)+{\cos}^2 {\left(\frac{\pi}{12}\right)}\cos2\pi \left(2q_1+q_2\right)+{\cos}^2 {\left(\frac{3\pi}{4}\right)}\cos2\pi \left(q_1+2q_2\right)-3\bigg] , \notag \\ 
E_{K_3}(\mathbf{q})&=\frac{1}{2}K_3\bigg[{\cos}^2 {\left(\frac{7\pi}{12}\right)}\cos4\pi \left(q_1+q_2\right)+{\cos}^2 {\left(\frac{\pi}{12}\right)}\cos4\pi q_2+{\cos}^2 {\left(\frac{3\pi}{4}\right)}\cos4\pi q_1-3\bigg] , \notag \\
E_{\Gamma_1}(\mathbf{q})&=\Gamma_1\bigg[\cos{\left(\frac{3\pi}{4}\right)}\cos(\frac{\pi}{12})\cos2\pi\left(q_1+q_2\right)+\cos{\left(\frac{7\pi}{12}\right)} \cos{\left(\frac{3\pi}{4}\right)}\cos2\pi q_2+\cos{\left(\frac{\pi}{12}\right)}{\cos{\left(\frac{7\pi}{12}\right)}}\cos2\pi q_1+3\bigg] , \notag \\
E_{\Gamma_2}(\mathbf{q})&=\Gamma_2\bigg[\cos{\left(\frac{3\pi}{4}\right)}\cos (\frac{\pi}{12})\cos2\pi \left(q_1-q_2\right)+\cos {\left(\frac{7\pi}{12}\right)}\cos {\left(\frac{3\pi}{4}\right)}\cos2\pi \left(2q_1+q_2\right) \notag \\
&\quad+\cos {\left(\frac{\pi}{12}\right)}{\cos {\left(\frac{7\pi}{12}\right)}}\left(q_1+2q_2\right)+3\bigg] , \notag \\
E_{\Gamma_3}(\mathbf{q})&=\Gamma_3\bigg[\cos{\left(\frac{3\pi}{4}\right)}\cos{\left(\frac{\pi}{12}\right)}\cos4\pi\left(q_1+q_2\right)+\cos{\left(\frac{7\pi}{12}\right)}\cos{\left(\frac{3\pi}{4}\right)}\cos4\pi q_2 \notag \\
&\quad+\cos{\left(\frac{\pi}{12}\right)}{\cos{\left(\frac{7\pi}{12}\right)}}\cos4\pi q_1+3\bigg] .
\end{align}

\vspace{2\baselineskip}
Based on Eq.~\eqref{EQ:3SPIN3SR}, the dispersion relations for the Heisenberg interaction of each nearest neighbor in the spin spiral method are expressed as:  
\begin{align}\label{eq:j1j2j3}
E_{J_1}(\mathbf{q})&= J_1\big[3-\cos2\pi q_1-\cos2\pi(q_1+q_2)-\cos(2\pi q_2)\big] , \notag \\ 
E_{J_2}(\mathbf{q})&= J_2\big[3-\cos2\pi(q_1+2q_2)-\cos2\pi(2q_1+q_2)
-\cos2\pi(q_1-q_2)\big] , \notag \\ 
E_{J_3}(\mathbf{q})&= J_3\big[3-\cos4\pi q_1-\cos4\pi(q_1+q_2)-\cos4\pi q_2\big] , \notag \\
E_{J_4}(\mathbf{q})&=\frac{1}{2}J_4\big[12-2\cos2\pi \left(2q_1+3q_2\right)-2\cos2\pi \left(q_1+3q_2\right)-2\cos2\pi \left(3q_1+2q_2\right)-2\cos2\pi (q_1-2q_2)  \notag \\
 &\quad-2\cos2\pi (3q_1+q_2)-2\cos2\pi (2q_1-q_2)\big] , \notag  \\
E_{J_5}(\mathbf{q})&=\frac{1}{2}J_5\big[6-\cos6\pi q_1-2cos{6}\pi(q_1+q_2)-2\cos6\pi q_2\big] ,\notag  \\ 
E_{J_6}(\mathbf{q})&=\frac{1}{2}J_6\big[6-2\cos4\pi \left(q_1+2q_2\right)-2\cos4\pi \left({2q}_1+q_2\right)-2\cos4\pi \left(q_1-q_2\right)\big] , \notag  \\
E_{J_7}(\mathbf{q})&=\frac{1}{2}J_7\big[12-2\cos2\pi \left(3q_1+4q_2\right)-2\cos2\pi \left(q_1+4q_2\right)-2\cos2\pi \left(4q_1+3q_2\right)-2\cos2\pi (q_1-3q_2) \notag  \\
&\quad-2\cos2\pi (4q_1+q_2)-2\cos2\pi (3q_1-q_2)\big] , \notag  \\
E_{J_8}(\mathbf{q})&=\frac{1}{2}J_8\big[6-\cos8\pi q_1-2\cos{8}\pi(q_1+q_2)-2\cos{8}\pi q_2\big] , \notag  \\
E_{J_9}(\mathbf{q})&=\frac{1}{2}J_9\big[12-\cos{2}\pi \left(2q_1+5q_2\right)-2\cos2\pi \left({3q}_1+5q_2\right)-2\cos2\pi \left(3q_1-2q_2\right)-2\cos2\pi ({5q}_1+3q_2) \notag  \\
&\quad-2\cos2\pi (5q_1+2q_2)-2\cos2\pi (2q_1-3q_2)\big]  , \notag  \\
E_{J_{10}}(\mathbf{q})&=\frac{1}{2}J_{10}\big[12-2\cos2\pi \left(q_1-4q_2\right)-2\cos2\pi \left({5q}_1+4q_2\right)-2\cos2\pi \left(5q_1+q_2\right)-2\cos2\pi ({4q}_1-q_2) \notag  \\
&\quad-2\cos2\pi (4q_1+5q_2)-2\cos2\pi (2q_1+5q_2)\big]  , \notag  \\
E_{J_{11}}(\mathbf{q})&=\frac{1}{2}J_{11}\big[6-\cos10\pi q_1-2\cos10\pi (q_1+q_2)-2\cos10\pi q_2\big] , \notag  \\
E_{J_{12}}(\mathbf{q})&=\frac{1}{2}J_{12}\big[6-2\cos6\pi \left(q_1-q_2\right)-2\cos6\pi \left({2q}_1+q_2\right)-2\cos6\pi \left(q_1+2q_2\right)\big] , \notag  \\
E_{J_{13}}(\mathbf{q})&=\frac{1}{2}J_{13}\big[12-2\cos2\pi \left(6q_1+2q_2\right)-2\cos2\pi \left(6q_1+4q_2\right)-2\cos2\pi \left(4q_1-2q_2\right)-2\cos2\pi (2q_1+6q_2) \notag  \\
&\quad-2\cos2\pi (4q_1+6q_2)-2\cos2\pi (2q_1-4q_2)\big]  , \notag  \\
E_{J_{14}}(\mathbf{q})&=\frac{1}{2}J_{14}\big[12-2\cos2\pi \left(6q_1+q_2\right)-2\cos2\pi \left(5q_1+6q_2\right)-2\cos2\pi \left(q_1-5q_2\right)-2\cos2\pi (q_1+6q_2) \notag  \\
&\quad-2\cos2\pi (6q_1+5q_2)-2\cos2\pi (5q_1-q_2)\big] , \notag  \\
E_{J_{15}}(\mathbf{q})&=\frac{1}{2}J_{15}\big[6-\cos12\pi q_1-2\cos{12}\pi (q_1+q_2)-2\cos{12}\pi q_2\big] .
\end{align}
Thereby, according to Eq.~\eqref{EQ:6Heisenberg interaction_S}, the interaction parameters of the Heisenberg interaction under the SOC alone were obtained (taking into account the third NN as well).
In case that SOC not considered, the system has only isotropic interactions. It is necessary to evaluate the $J_{Ni}$ for more distant nearest neighbors by considering the long-range magnetic sequence of the system. 
Then, the Heisenberg interaction is considered to the fifteen NN according to Eq.~\eqref{Eq:9NHeisenberg interaction} .

\setcounter{equation}{0} 
\section{MAE of individual parameters under different magnetic order} \label{Appendix B}
Under ZZ order, the Kitaev interactions are listed as :
\begin{align}
E_{K_1}&=K_1\left[-{F\left(\theta,\phi_a\right)}^2+\frac{1}{4}\right] , \notag \\
E_{K_2}&=K_2\left[{F\left(\theta,\phi_a\right)}^2-\frac{1}{4}\right] , \notag \\
E_{K_3}&=K_3\left[-{F\left(\theta,\phi_b\right)}^2+{F\left(\theta,\phi_a\right)}^2-{F\left(\theta,\phi_c\right)}^2+\frac{1}{4}\right] , \notag \\
E_{\Gamma_1}&={2\Gamma}_1\left[-F\left(\theta,\phi_c\right)F\left(\theta,\phi_b\right)+\frac{1}{4}\right] , \notag \\
E_{\Gamma_2}&={2\Gamma}_2\left[F\left(\theta,\phi_c\right)F\left(\theta,\phi_b\right)-\frac{1}{4}\right] , \notag \\
E_{\Gamma_3}& = 2\Gamma_2\left[F\left(\theta,\phi_c\right)F\left(\theta,\phi_b\right)-F\left(\theta,\phi_a\right)F\left(\theta,\phi_c\right)-F\left(\theta,\phi_b\right)F\left(\theta,\phi_a\right)+\frac{1}{4}\right],
\end{align}
with $F=\sqrt{\frac{2}{3}}\sin{\theta}\cos{\phi_a}+\sqrt{\frac{1}{3}}\cos{\theta}$. The $\phi_a$, $\phi_b$, and $\phi_c$ are $\phi_a=\phi-\frac{5\pi}{6}$, $\phi_b=\phi_a+\frac{2\pi}{3}$, and $\phi_c=\phi_a-\frac{2\pi}{3}$.
\end{widetext}

%

\clearpage

\onecolumngrid

\newpage

\newcounter{sectionSM}
\newcounter{equationSM}
\newcounter{figureSM}
\newcounter{tableSM}
\stepcounter{equationSM}
\setcounter{section}{0}
\setcounter{equation}{0}
\setcounter{figure}{0}
\setcounter{table}{0}
\setcounter{page}{1}
\makeatletter
\renewcommand{\thesection}{\textsc{S}\arabic{section}}
\renewcommand{\theequation}{\textsc{S}\arabic{equation}}
\renewcommand{\thefigure}{\textsc{S}\arabic{figure}}
\renewcommand{\thetable}{\textsc{S}\arabic{table}}


\begin{center}
{\large{\bf Supplemental Material for\\
``Evidence of Kitaev interaction in the monolayer 1T-CrTe$_2$''}}
\end{center}
\begin{center}
Can Huang$^{1,\;2}$, Bingjie Liu$^{1,\;2}$, Lingzi Jiang$^{1,\;2}$, Yanfei Pan$^{1,\;2}$, Jiyu Fan$^{1,\;2}$, Daning Shi$^{1,\;2}$, Chunlan Ma$^3$, Qiang Luo$^{1,\;2}$, Yan Zhu$^{1,\;2}$ \\
\quad\\
$^1$\textit{College of Science, Nanjing University of Aeronautics and Astronautics, Nanjing, 211106, China}\\
$^2$\textit{Key Laboratory of Aerospace Information Materials and Physics, MIIT, Nanjing, 211106, China} \\
$^3$\textit{Jiangsu Key Laboratory of Micro and Nano Heat Fluid Flow Technology and Energy Application, School of Mathematics and Physics, Suzhou University of Science and Technology, Suzhou 215009, China}\\
(Dated: May 10, 2023)
\quad\\
\end{center}

\vspace{-0.00cm}
\section{HSE06 band structures of monolayer 1T-CrTe$_2$}
In order to evaluate the band renormalization effect, we provide the HSE06 energy band structures without and with spin-orbit coupling (SOC). As shown in Fig. S1(a), the HSE06 band structures is basically consistent with the PBE+$U$ band without SOC. By adding SOC, the monolayer 1T-CrTe$_2$ remains metallic from the HSE06 electronic band structures.  A small separation between the energy band profiles of HSE06 and PBE+$U$ near the Fermi energy level is observed in Fig. S1(b). 
Importantly, the magnetization angle calculated by using the PBE+$U$ energy band structure is in agreement with experiment~\cite{smXian2022}. Furthermore, we have checked the correctness of the PBE+$U$ energy band diagrams against previous work, for which the energy band ($U$ = 2 eV) diagrams are practically the same~\cite{smLi2021}. Thus, in the case of 1T-CrTe$_2$, the energy band renormalization is ignored and the strong correlation effect partially occupying the Cr $d$ state is corrected by the PBE+$U$ used in the main text. 
\begin{figure}[!ht]
\centering
\includegraphics[width=0.65\columnwidth, clip]{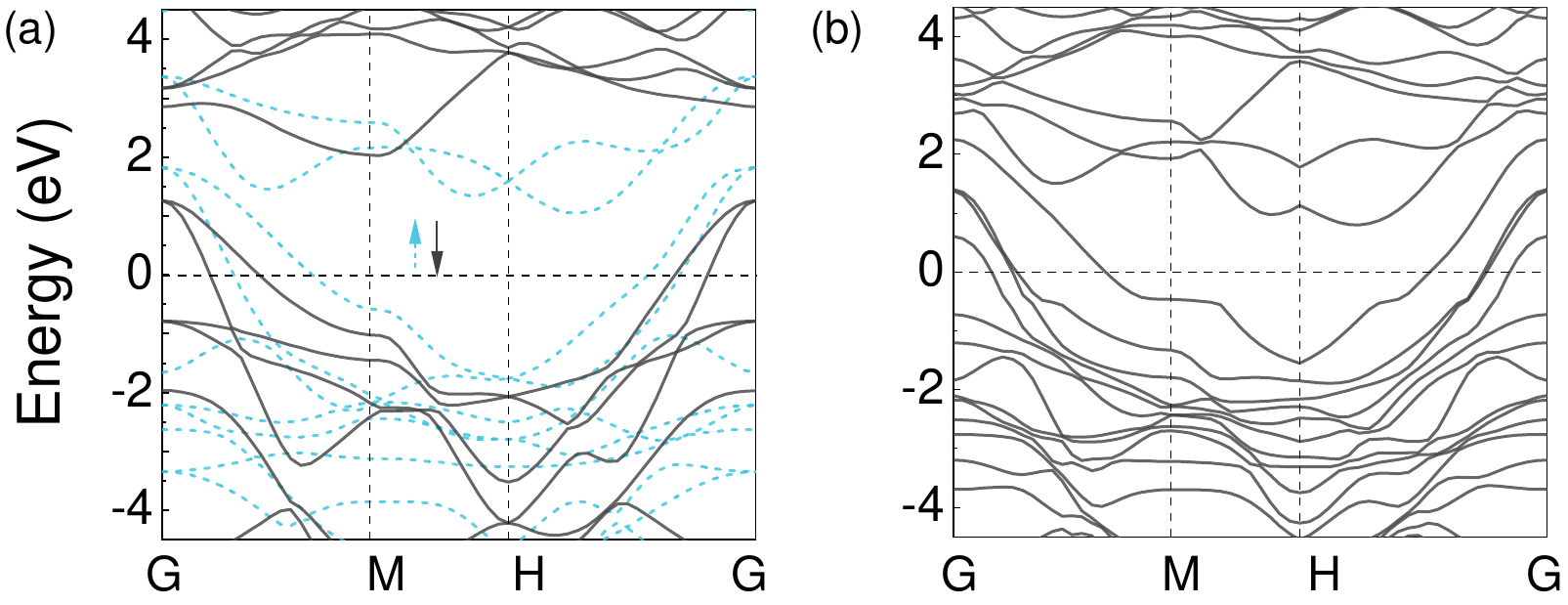}\\
\caption{ (a) and (b) are the HSE06 electronic band structures without and with SOC, respectively, where blue represents spin-up and dark represents spin-down.
  }\label{FIG-s1}
\end{figure}


\section{Calculations of the monolayer RuCl$_3$}
We are presently utilizing density-functional theory (DFT) together with the spin spiral method to investigate monolayer RuCl$_3$. We select two well-studied models, $J_1$-$K$-$\Gamma$-$\Gamma'$ and $J_1$-$J_3$-$K$-$\Gamma$, as illustrated in the Fig. S2. The interaction parameters of the different models are then derived from Eq.~\eqref{EQ:1Ham} and Eq.~\eqref{EQ:2Ham} in association with the generalised Bloch conditions. SOC is included throughout the calculation process. We added Hubbard $U$ parameter of 1.5 eV in order to treat the strong on-site Coulomb interaction of localized $d$ electrons of the Ru atoms. K-point grids of $11\times11\times1$ were used within the Monkhorst-Park scheme. 
 We calculat the spin-spiral dispersion energy of monolayer RuCl$_3$. The best-fit parameters for the $J_1$-$K$-$\Gamma$-$\Gamma'$ model give  Heisenberg interactions $J_1$ = 1.68 meV, $K = -10.9$ meV, $\Gamma$ = 3.29 meV, and $\Gamma' = -0.28$ meV. 
 As for the $J_1$-$J_3$-$K$-$\Gamma$ model, we find that $J_1$ = 0.36 meV, $J_3 = -0.27$ meV, $K = -11.0$ meV, and $\Gamma$ = 5.30 meV.
Note that the positivity and negativity of the $J$ values are determined here based on Eq.~\eqref{EQ:1Ham}, opposite to the sign of the conventional Heisenberg interaction formula.
In both cases, the dominance of $K$ and $\Gamma$ as well
as the main features of negative $K$ and positive $\Gamma$ are
preserved, which is in accordance with the consensus in
the community. Moreover, the negative $\Gamma'$ or positive $J_3$ which is equivalent to the conventional positive Heisenberg $J$ can stabilize the zigzag ordering of the RuCl$_3$.
\begin{figure}[!ht]
\centering
\includegraphics[width=0.4\columnwidth, clip]{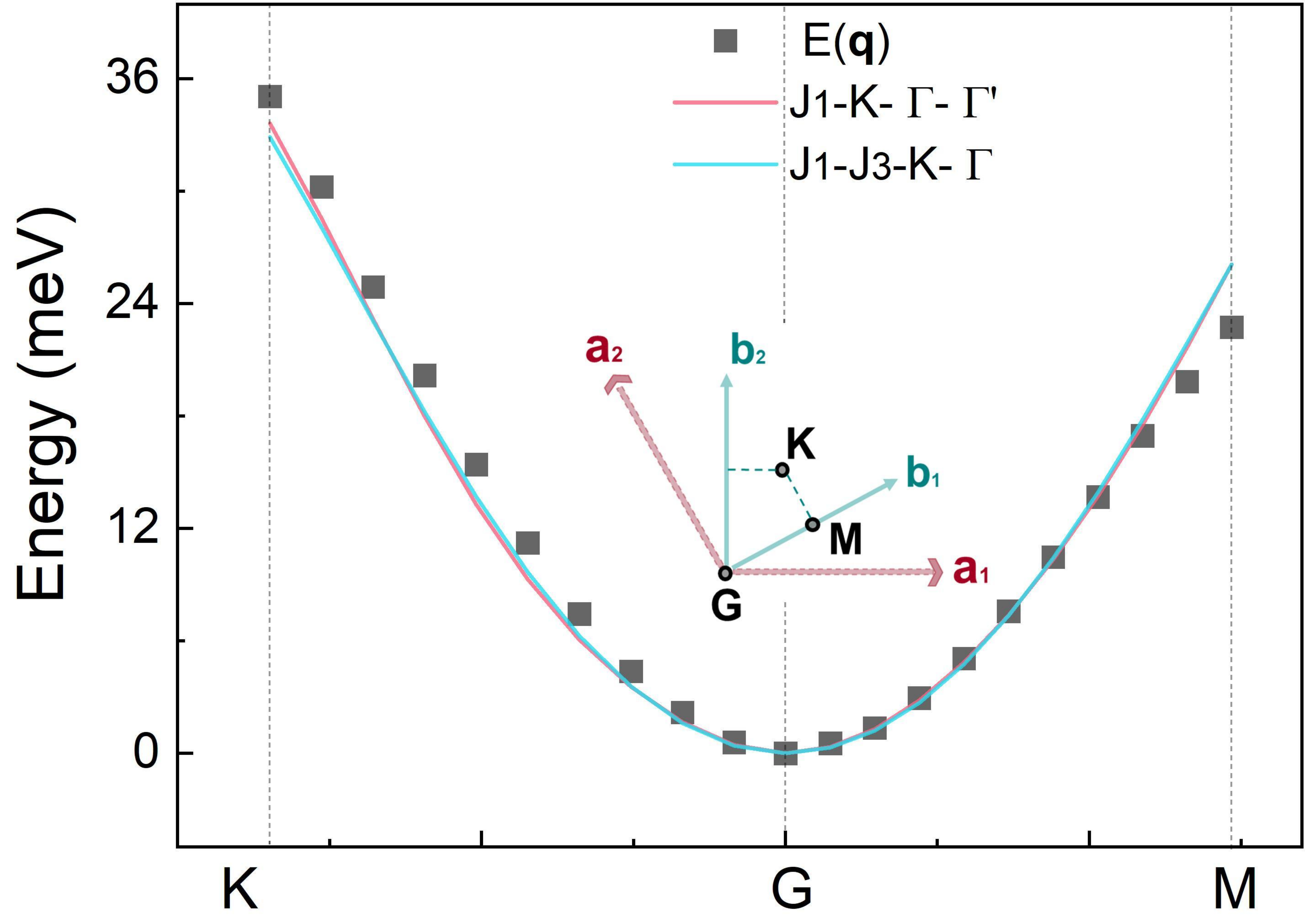}\\
\caption{ $E(\mathbf{q})$ as a function of the spiral wave vector $\mathbf{q}$ in the RuCl$_3$ monolayer. The black squares are the practical calculated values and the pink and blue curves are obtained by fitting the parameters using the least-squares method. The pink represents the $J_1$-$K$-$\Gamma$-$\Gamma'$ model and while the blue represents the $J_1$-$J_3$-$K$-$\Gamma$ model.
  }\label{FIG-s2}
\end{figure}
 

\vspace{4.00cm}
\section{Magnetic moment diagram for different $E(\mathbf{q})$}
As shown in Fig. S3, the magnetic moment diagrams for the high symmetry points and the lowest energy points of the system in the spin-spiral dispersion relation are presented.
\begin{figure}[!ht]
\centering
\includegraphics[width=0.6\columnwidth, clip]{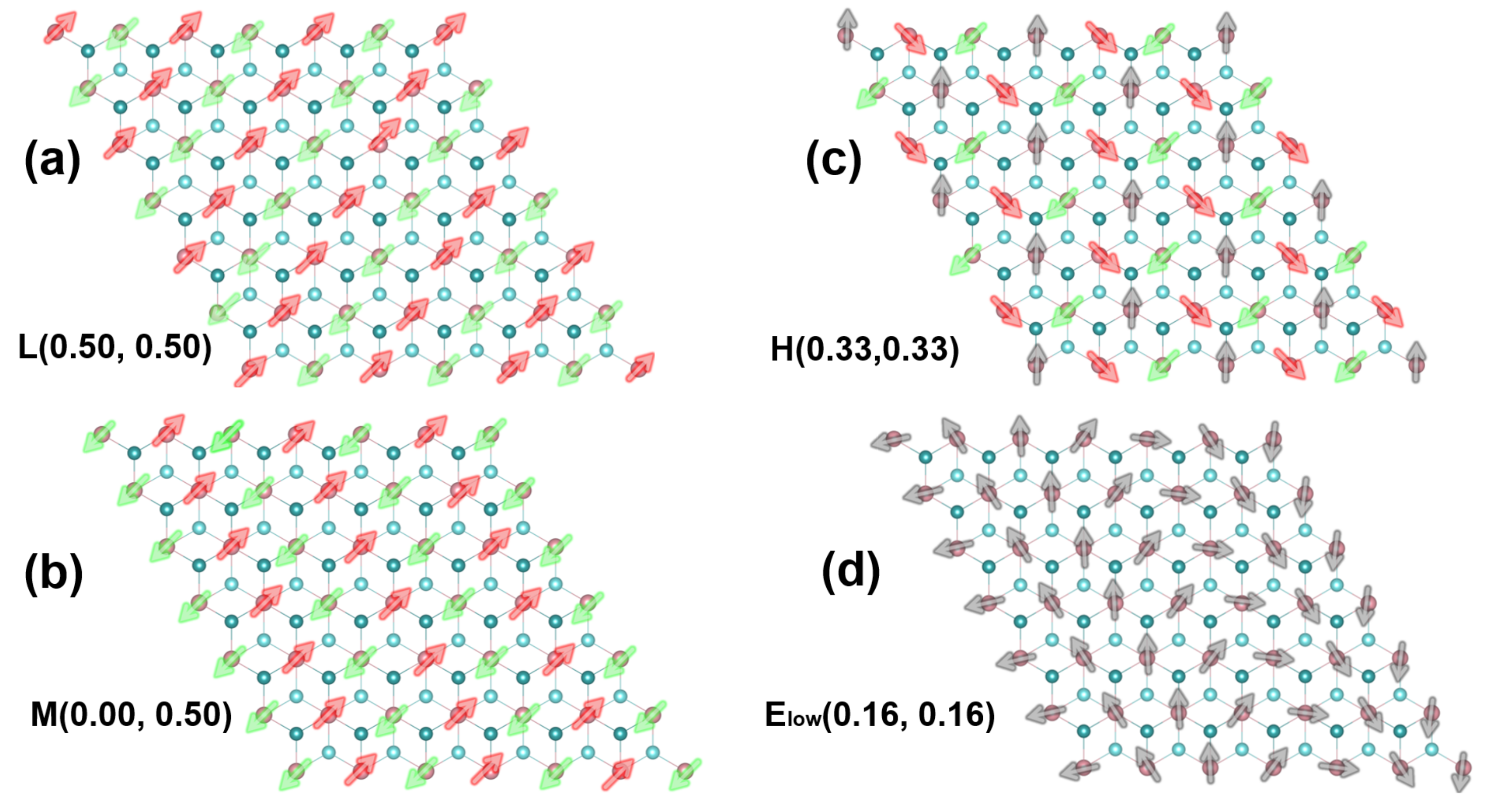 }\\
\caption{The magnetic moment maps that correspond to the high symmetry and the lowest energy point in the $E_S(\mathbf{q})$ based on the Generalized Bloch equation in Eq.~\eqref{EQ:3SPIN3SR}, respectively. 
(a-d) The magnetic moment maps corresponding to the high symmetry sites $\mathbf{L}$, $\mathbf{M}$, $\mathbf{H}$, and the minimum energy of $E_{low}$ in the Brillouin zone, respectively. 
The magnetic moments are in the $\textbf{ij}$ plane, and the color presents the different directions.
  }\label{FIG-s3}
\end{figure}

\vspace{2.00cm}
\section{Determination of the Effective Model}
Without term $J$, only the Kitaev and SIA terms are considered, as in Fig. S4. 
The the fitted curves also deviated significantly from the dispersion relation $E_S(\mathbf{q})$ as shown in Fig. S4(a).
Meanwhile, it is clearly shown that the simplest MAE-FM cannot be fitted as shown in Fig. S4(b).
These indicate that contribution of Heisenberg term in the metallic 1T-CrTe$_2$ remains very large.
\begin{figure}[!ht]
\centering
\includegraphics[width=0.85\columnwidth, clip]{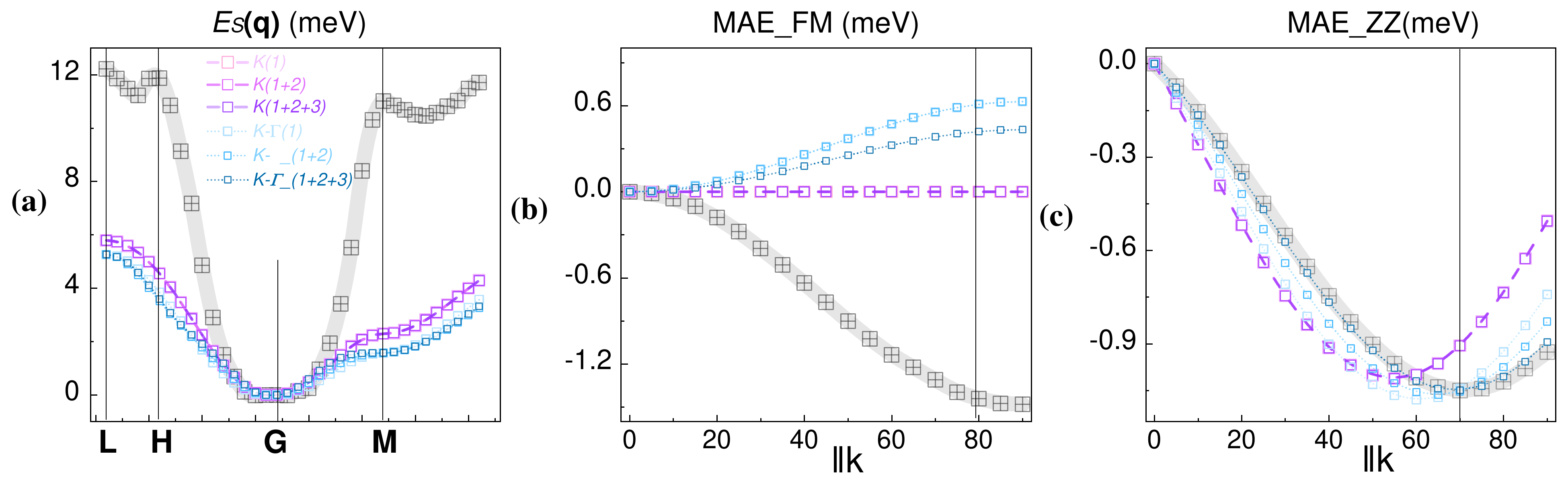}\\
\caption{Without considering $J$, the other parameters are considered for the cases of N, NN, and NNN fitted separately.
(a-c) The fitted graphs of $E_S(\mathbf{q})$, MAE-FM, MAE-ZZ are shown in sequence.
Each model is represented by a unique color family, while the different nearest neighbors are shown by different shades of the same color family. 
The $K$-$A_\mathbf{k}$ models are represented by a purple color family and $K$-$\Gamma$-$A_\mathbf{k}$ by a blue color family.
  }\label{FIG-s4}
\end{figure}

\vspace{1cm}
Most remarkably, in the Fig. S5, when fitting MAE-ZZ singularly, the fitted parameters will seriously deviate from the calculated values of $E_S(\mathbf{q})$ and MAE-FM. Moreover, without considering the SIA, there is also a bias in the single-fit MAE-ZZ. 
\begin{figure}[!ht]
\centering
\includegraphics[height=6cm, clip]{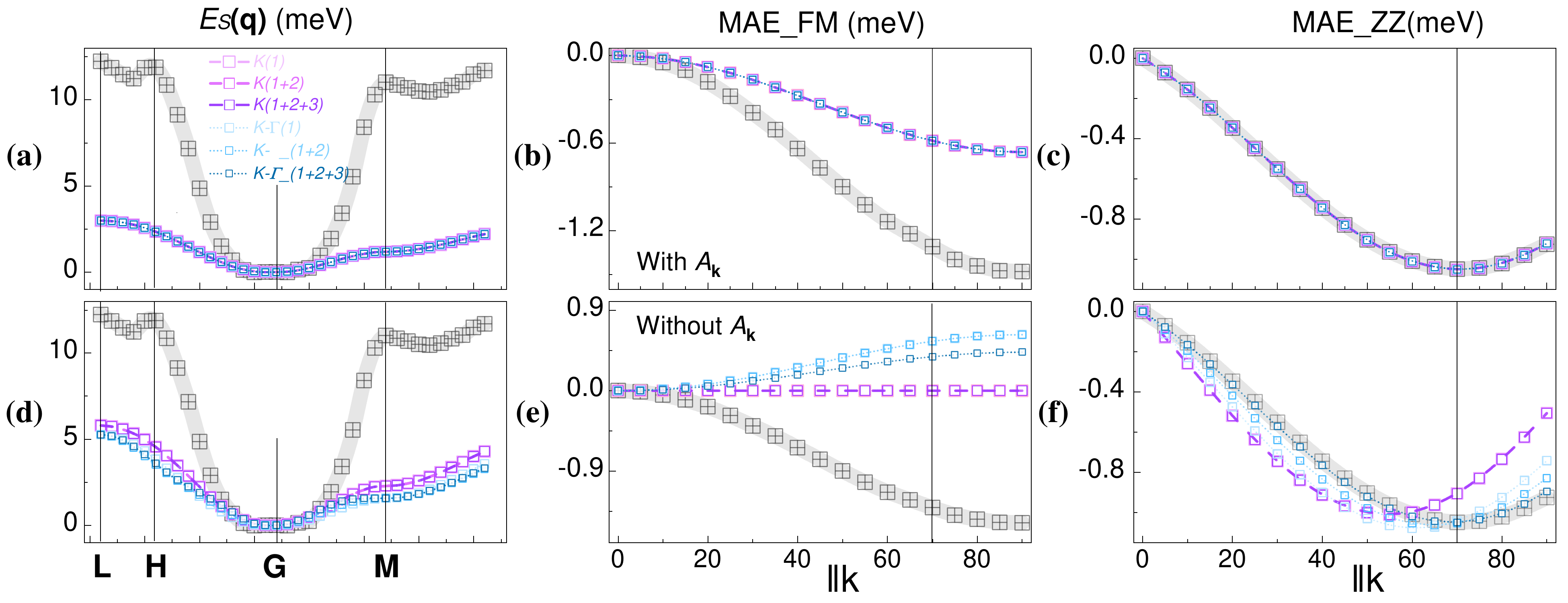}\\
\caption{A single fitted to the ZZ data with and without SIA separately. 
(a-c) Models with different Kitaev and $A_\mathbf{k}$ term for a single fitted MAE-ZZ data to obtain the interaction parameters, and fitted $E_s(\mathbf{q})$, MAE-FM, and MAE-ZZ, respectively.
(d-f) Figures for the fitted results without the $A_\mathbf{k}$ term in the model.
  }\label{FIG-s5}
\end{figure}

%


\begin{thebibliography}{99}%

\bibitem{Wen2019npj}
  J. Wen, S. L. Yu, S. Li, W. Yu, and J. X. Li, 
 Experimental identification of quantum spin liquids,
 \href{https://doi.org/10.1038/s41535-019-0151-6}{npj Quantum Mater. \textbf{4}, 12 (2019)}. 

\bibitem{Rousochatzakis2023}
 I. Rousochatzakis, N. B. Perkins, Q. Luo, and H. Y. Kee,
  Beyond Kitaev physics in strong spin-orbit coupled magnets,
 \href{https://doi.org/10.48550/arXiv.2308.01943}{arXiv: 2308.01943 (2023)}.
 
\bibitem{Tokura2017}
 Y. Tokura, M. Kawasaki, and N. Nagaosa,
  Emergent functions of quantum materials,
 \href{ https://doi.org/10.1038/nphys4274}{Nat. Phys. \textbf{13}, 1056–1068 (2017)}.

\bibitem{Trebst2021}
 S. Trebst and C. Hickey, 
 Kitaev materials,
 \href{ https://doi.org/10.1016/j.physrep.2021.11.003}{Phys. Rep. \textbf{950}, 1–37 (2022)}.

\bibitem{Kitaev2006}
 A. Kitaev,
 Anyons in an exactly solved model and beyond,
 \href{https://www.sciencedirect.com/science/article/pii/S0003491605002381}{Ann. Phys. \textbf{321}, 2-111 (2006)}.

 \bibitem{Jackeli2009}
 G. Jackeli and G. Khaliullin,
 Mott insulators in the strong spin-orbit coupling limit: from Heisenberg to a quantum compass and Kitaev models,
 \href{https://link.aps.org/doi/10.1103/PhysRevLett.102.017205}{Phys. Rev. Lett. \textbf{102}, 017205 (2009)}.

\bibitem{Plumb2014}
K. W. Plumb, J. P. Clancy, L. J. Sandilands, V. V. Shankar, Y. F. Hu, K. S. Burch, H. Y. Kee, and Y. J. Kim, 
 $\alpha$-RuCl$_3$: A spin-orbit assisted Mott insulator on a honeycomb lattice,
 \href{https://link.aps.org/doi/10.1103/PhysRevB.90.041112}{Phys. Rev. B \textbf{90}, 041112(R) (2014)}.

 \bibitem{Banerjee2016}
A. Banerjee, C. Bridges, J.-Q. Yan, A. Aczel, L. Li, M. Stone, G. Granroth, M. Lumsden, Y. and Yiu, J. Knolle,
 Proximate Kitaev quantum spin liquid behaviour in a honeycomb magnet,
 \href{https://doi.org/10.1038/nmat4604}{Nat. Mater. \textbf{15}, 733–740 (2016)}.

  \bibitem{Ran2017}
K. Ran, J. Wang, W. Wang, Z.-Y. Dong, X. Ren, S. Bao, S. Li, Z. Ma, Y. Gan, Y. Zhang, J.T. Park, G. Deng, S. Danilkin, S.-L. Yu, J.-X. Li, and J. Wen, 
 Spin-Wave Excitations Evidencing the Kitaev Interaction in Single Crystalline $\alpha$-RuCl$_3$,
 \href{https://link.aps.org/doi/10.1103/PhysRevLett.118.107203}{Phys. Rev. Lett. \textbf{118}, 107203 (2017)}.

  \bibitem{Lee2020}
 I. Lee, F. G. Utermohlen, D. Weber, K. Hwang, C. Zhang, J. van Tol, J. E. Goldberger, N. Trivedi, and P. C. Hammel, 
Fundamental spin interactions underlying the magnetic anisotropy in the Kitaev ferromagnet ${\mathrm{CrI}}_{3}$,
 \href{https://link.aps.org/doi/10.1103/PhysRevLett.124.017201}{Phys. Rev. Lett. \textbf{124}, 017201 (2020)}.

\bibitem{Cai2021prb}
 Z. Cai, S. Bao, Z.-L. Gu, Y.-P. Gao, Z. Ma, Y. Shangguan, W. Si, Z.-Y. Dong, W. Wang, Y. Wu et al, 
Topological magnon insulator spin excitations in the two-dimensional ferromagnet ${\mathrm{CrBr}}_{3}$,
 \href{https://link.aps.org/doi/10.1103/PhysRevB.104.L020402}{Phys. Rev. B \textbf{104}, L020402 (2021)}. 

\bibitem{Xu2018}
 C. Xu, J. Feng, H. Xiang, and L. Bellaiche, 
 Interplay between Kitaev interaction and single ion anisotropy in ferromagnetic CrI$_3$ and CrGeTe$_3$ monolayers,
 \href{https://doi.org/10.1038/s41524-018-0115-6}{npj Comput. Mater. \textbf{4}, 57 (2018)}.

 \bibitem{Stavropoulos2019}
 P. P. Stavropoulos, D. Pereira, and H.-Y. Kee, 
 Microscopic Mechanism for a Higher-Spin Kitaev Model,
 \href{https://link.aps.org/doi/10.1103/PhysRevLett.123.037203}{Phys. Rev. Lett. \textbf{123}, 037203 (2019)}.

  \bibitem{Xu2020}
 C. Xu, J. Feng, M. Kawamura, Y. Yamaji, Y. Nahas, S. Prokhorenko, Y. Qi, H. Xiang, and L. Bellaiche, 
Possible Kitaev Quantum Spin Liquid State in 2D Materials with $S=3/2$,
 \href{https://link.aps.org/doi/10.1103/PhysRevLett.124.087205}{Phys. Rev. Lett. \textbf{124}, 087205 (2020)}.
 
 \bibitem{Jaeschke-Ubiergo2021prb}
  R. Jaeschke-Ubiergo, E. Su\'arez Morell, and A. S. Nunez, 
 Theory of magnetism in the van der Waals magnet ${\mathrm{CrI}}_{3}$,
 \href{https://link.aps.org/doi/10.1103/PhysRevB.103.174410}{Phys. Rev. B \textbf{103}, 174410 (2021)}. 

 \bibitem{Freitas2015}
D. C. Freitas, R. Weht, A. Sulpice, G. Remenyi, P. Strobel, F. Gay, J. Marcus, and M. N\'{u}\~{n}ez-Regueiro,  
Ferromagnetism in layered metastable 1T-CrTe$_2$,
 \href{https://dx.doi.org/10.1088/0953-8984/27/17/176002}{J. Phys.: Condens. Matter \textbf{27}, 176002 (2015)}.

 \bibitem{Zhang2021}
X. Zhang, Q. Lu, W. Liu, W. Niu, J. Sun, J. Cook, M. Vaninger, P. F. Miceli, D. J. Singh, and S.-W. Lian,  
Room-temperature intrinsic ferromagnetism in epitaxial CrTe$_2$ ultrathin films,
 \href{https://doi.org/10.1038/s41467-021-22777-x}{Nat. Commun. \textbf{12}, 2492 (2021)}.

 \bibitem{Meng2021}
 L. Meng, Z. Zhou, M. Xu, S. Yang, K. Si, L. Liu, X. Wang, H. Jiang, B. Li, P. Qin, P. Zhang, J. Wang, Z. Liu, P. Tang, Y. Ye, W. Zhou, L. Bao, H.-J. Gao, and Y. Gong,
 Anomalous thickness dependence of Curie temperature in air-stable two-dimensional ferromagnetic 1T-CrTe$_2$ grown by chemical vapor deposition,
 \href{https://doi.org/10.1038/s41467-021-21072-z}{Nat. Commun. \textbf{12}, 809 (2021)}.
 
 \bibitem{Sun2021}
Y. Sun, P. Yan, J. Ning, X. Zhang, Y. Zhao, Q. Gao, M. Kanagaraj, K. Zhang, J. Li, and X. Lu,
Ferromagnetism in layered metastable 1T-CrTe$_2$,
 \href{https://doi.org/10.1063/5.0041531}{ AIP Adv. \textbf{11}, 035138 (2021)}.

 \bibitem{Xian2022}
J.-J. Xian, C. Wang, J.-H. Nie, R. Li, M. Han, J. Lin, W.-H. Zhang, Z.-Y. Liu, Z.-M. Zhang, M.-P. Miao, Y. Yi, S. Wu, X. Chen, J. Han, Z. Xia, W. Ji, and Y.-S. Fu,  Spin mapping of intralayer antiferromagnetism and field-induced spin reorientation in monolayer CrTe$_2$,
 \href{https://doi.org/10.1038/s41467-021-27834-z}{ Nat. Commun. \textbf{13}, 257 (2022)}.

 \bibitem{Chaloupka2015}
 J. Chaloupka and G. Khaliullin, 
Hidden symmetries of the extended Kitaev-Heisenberg model: Implications for the honeycomb-lattice iridates ${A}_{2}{\mathrm{IrO}}_{3}$,
 \href{https://link.aps.org/doi/10.1103/PhysRevB.92.024413}{Phys. Rev. B \textbf{92}, 024413 (2015)}.

\bibitem{Sizyuk2014}
 Y. Sizyuk, C. Price, P. W\"olfle, and N. B. Perkins,
Importance of anisotropic exchange interactions in honeycomb iridates: Minimal model for zigzag antiferromagnetic order in ${\mathrm{Na}}_{2}{\mathrm{IrO}}_{3}$,
 \href{https://link.aps.org/doi/10.1103/PhysRevB.90.155126}{Phys. Rev. B \textbf{90}, 155126 (2014)}.
 
\bibitem{Winter_2017}
  S. M. Winter,  A. A. Tsirlin, M. Daghofer, J. van den Brink, Y. Singh, P. Gegenwart, and R. Valent\'{\i}, 
Models and materials for generalized Kitaev magnetism,
 \href{https://dx.doi.org/10.1088/1361-648X/aa8cf5}{ J. Phys.: Condens. Matter \textbf{29}, 493002 (2017)}.

 \bibitem{Lv2015PRB}
 H. Y. Lv, W. J. Lu, D. F. Shao, Y. Liu, and Y. P. Sun, 
 Strain-controlled switch between ferromagnetism and antiferromagnetism in $1T\text{\ensuremath{-}}\mathrm{Cr}{X}_{2}$ ($X=\text{Se}$, Te) monolayers,
 \href{https://link.aps.org/doi/10.1103/PhysRevB.92.214419}{Phys. Rev. B \textbf{92}, 214419 (2015)}.

\bibitem{Li2015Sci}
 Y. Li, H. Liao, Z. Zhang, S. Li, F. Jin, L. Ling, L. Zhang, Y. Zou, L. Pi, Z. Yang, J. Wang, Z. Wu, Q. Zhang, 
  Gapless quantum spin liquid ground state in the two-dimensional spin$-1/2$ triangular antiferromagnet YbMgGaO$_4$,
 \href{https://doi.org/10.1038/srep16419}{Sci. Rep. \textbf{5}, 16419 (2015)}.
 
\bibitem{Li2015}
 Y. Li, G. Chen, W. Tong, L. Pi, J. Liu, Z. Yang, X. Wang, and Q. Zhang, 
Rare-earth triangular lattice spin liquid: a single-crystal study of YbMgGaO$_4$,
 \href{https://link.aps.org/doi/10.1103/PhysRevLett.115.167203}{Phys. Rev. Lett. \textbf{115}, 167203 (2015)}.

 \bibitem{Shen2016}
Y. Shen, Y.-D. Li, H. Wo, Y. Li, S. Shen, B. Pan, Q. Wang, H. Walker, P. Steffens, and M. Boehm, 
Evidence for a spinon Fermi surface in a triangular-lattice quantum-spin-liquid candidate,
 \href{https://doi.org/10.1038/nature20614}{Nature \textbf{540}, 559 (2016)}.

\bibitem{Paddison2017}
J. A. Paddison, M. Daum, Z. Dun, G. Ehlers, Y. Liu, M. B. Stone, H. Zhou, and M. Mourigal, 
Continuous excitations of the triangular-lattice quantum spin liquid YbMgGaO$_4$,
 \href{https://doi.org/10.1038/nphys3971}{Nat. Phys. \textbf{13}, 117-122 (2017)}.

\bibitem{Bordelon2019}
M. M. Bordelon, E. Kenney, C. Liu, T. Hogan, L. Posthuma, M. Kavand, Y. Lyu, M. Sherwin, N. P. Butch, and C. Brown, 
Field-tunable quantum disordered ground state in the triangular-lattice antiferromagnet NaYbO$_2$,
 \href{https://doi.org/10.1038/s41567-019-0594-5}{Nat. Phys. \textbf{15}, 1058 (2019)}. 

\bibitem{Ding2019}
 L. Ding, P. Manuel, S. Bachus, F. Gru\ss{}ler, P. Gegenwart, J. Singleton, R. D. Johnson, H. C. Walker, D. T. Adroja, and A. D. Hillier, 
Gapless spin-liquid state in the structurally disorder-free triangular antiferromagnet NaYbO$_2$,
 \href{https://link.aps.org/doi/10.1103/PhysRevB.100.144432}{Phys. Rev. B \textbf{100}, 144432 (2019)}. 

\bibitem{Arh2022}
T. Arh, B. Sana, M. Pregelj, P. Khuntia, Z. Jagli\v{c}i\'c, M. Le, P. Biswas, P. Manuel, L. Mangin-Thro, and A. Ozarowski, 
The Ising triangular-lattice antiferromagnet neodymium heptatantalate as a quantum spin liquid candidate,
 \href{https://doi.org/10.1038/s41563-021-01169-y}{Nat. Mater. \textbf{21}, 416–422 (2022)}. 

\bibitem{Liu2018}
W. Liu, Z. Zhang, J. Ji, Y. Liu, J. Li, X. Wang, H. Lei, G. Chen, and Q. Zhang, 
Rare-earth chalcogenides: a large family of triangular lattice spin liquid candidates,
 \href{https://dx.doi.org/10.1088/0256-307X/35/11/117501}{Chin. Phys. Lett. \textbf{35}, 117501 (2018)}.  

\bibitem{Ortiz2023}
 B. R. Ortiz, P. M. Sarte, A. H. Avidor, A. Hay, E. Kenney, A. I. Kolesnikov, D. M. Pajerowski, A. A. Aczel, K. M. Taddei and C. M. Brown, 
Quantum disordered ground state in the triangular-lattice magnet NaRuO$_2$,
 \href{https://doi.org/10.1038/s41567-023-02039-x}{Nat. Phys. 1-7 (2023)}.   
  
\bibitem{Li2015NewJ}
K. Li, S.-L. Yu, and J.-X. Li, 
Global phase diagram, possible chiral spin liquid, and topological superconductivity in the triangular Kitaev–Heisenberg model,
 \href{https://dx.doi.org/10.1088/1367-2630/17/4/043032}{New J. Phys. \textbf{17}, 043032 (2015)}. 

\bibitem{Jackeli2015}
 G. Jackeli and A. Avella, 
Quantum order by disorder in the Kitaev model on a triangular lattice,
 \href{https://link.aps.org/doi/10.1103/PhysRevB.92.184416}{Phys. Rev. B \textbf{92}, 184416 (2015)}.  

\bibitem{Becker2015}
 M. Becker, M. Hermanns, B. Bauer, M. Garst, and S. Trebst,
Spin-orbit physics of $j=\frac{1}{2}$ Mott insulators on the triangular lattice,
 \href{https://link.aps.org/doi/10.1103/PhysRevB.91.155135}{Phys. Rev. B \textbf{91}, 155135 (2015)}.

\bibitem{Maksimov2019}
P. A. Maksimov, Z. Zhu, S. R. White, and A. L. Chernyshev,
Anisotropic-exchange magnets on a triangular lattice: spin waves, accidental degeneracies, and dual spin liquids,
 \href{https://link.aps.org/doi/10.1103/PhysRevX.9.021017}{Phys. Rev. X \textbf{9}, 021017 (2019)}.

\bibitem{Luo2017}
Q. Luo, S. Hu, B. Xi, J. Zhao, and X. Wang, 
Ground-state phase diagram of an anisotropic spin-$\frac{1}{2}$ model on the triangular lattice,
 \href{https://link.aps.org/doi/10.1103/PhysRevB.95.165110}{Phys. Rev. B \textbf{95}, 165110 (2017)}. 

 \bibitem{Wang2021prb}
S. Wang, Z. Qi, B. Xi, W. Wang, S. L. Yu, and J. X. Li, 
Comprehensive study of the global phase diagram of the $J\ensuremath{-}K\ensuremath{-}\mathrm{\ensuremath{\Gamma}}$ model on a triangular lattice,
 \href{https://link.aps.org/doi/10.1103/PhysRevB.103.054410}{Phys. Rev. B \textbf{103}, 054410 (2021)}. 

\bibitem{Li2021}
S. Li, S.-S. Wang, B. Tai, W. Wu, B. Xiang, X.-L. Sheng, and S. A. Yang, 
Tunable anomalous Hall transport in bulk and two-dimensional $1T\ensuremath{-}{\mathrm{CrTe}}_{2}$: A first-principles study,
 \href{https://link.aps.org/doi/10.1103/PhysRevB.103.045114}{Phys. Rev. B \textbf{103}, 045114 (2021)}. 

\bibitem{Wu2022}
L. Wu, L. Zhou, X. Zhou, C. Wang, and W. Ji, 
In-plane epitaxy-strain-tuning intralayer and interlayer magnetic coupling in ${\mathrm{CrSe}}_{2}$ and ${\mathrm{CrTe}}_{2}$ monolayers and bilayers,
 \href{https://link.aps.org/doi/10.1103/PhysRevB.106.L081401}{Phys. Rev. B \textbf{106}, L081401 (2022)}.

\bibitem{Aghaee2022}
A. Karbalaee Aghaee, S. Belbasi, and H. Hadipour, 
Ab initio calculation of the effective Coulomb interactions in $M{X}_{2} (M=\mathrm{Ti},\mathrm{V},\mathrm{Cr},\mathrm{Mn},\mathrm{Fe},\mathrm{Co},\mathrm{Ni};X=\mathrm{S},\mathrm{Se},\mathrm{Te})$: Intrinsic magnetic ordering and Mott phase,
 \href{https://link.aps.org/doi/10.1103/PhysRevB.105.115115}{Phys. Rev. B \textbf{105}, 115115 (2022)}.  

\bibitem{Liu2022}
Y. Liu, S. Kwon, G. J. de Coster, R. K. Lake, and M. R. Neupane, 
Structural, electronic, and magnetic properties of ${\mathrm{CrTe}}_{2}$,
 \href{https://link.aps.org/doi/10.1103/PhysRevMaterials.6.084004}{Phys. Rev. Mater. \textbf{6}, 084004 (2022)}. 

\bibitem{Rau2014}
J. G. Rau, E. K.-H. Lee, and H.-Y. Kee,
Generic Spin Model for the Honeycomb Iridates beyond the Kitaev Limit,
 \href{https://link.aps.org/doi/10.1103/PhysRevLett.112.077204}{Phys. Rev. Lett. \textbf{112}, 077204 (2014)}. 

 \bibitem{Rousochatzakis2017}
I. Rousochatzakis and N. B. Perkins,
Classical Spin Liquid Instability Driven By Off-Diagonal Exchange in Strong Spin-Orbit Magnets,
 \href{https://link.aps.org/doi/10.1103/PhysRevLett.118.147204}{Phys. Rev. Lett. \textbf{118}, 147204 (2017)}. 

\bibitem{Kresse1999}
G. Kresse and D. Joubert,
From ultrasoft pseudopotentials to the projector augmented-wave method,
 \href{https://link.aps.org/doi/10.1103/PhysRevB.59.1758}{Phys. Rev. B \textbf{59}, 1758 (1999)}.  

\bibitem{Perdew1996}
J. P. Perdew, K. Burke, and M. Ernzerhof, 
Generalized Gradient Approximation Made Simple,
 \href{https://link.aps.org/doi/10.1103/PhysRevLett.77.3865}{Phys. Rev. Lett. \textbf{77}, 3865 (1996)}. 

\bibitem{Hummer2009}
K. Hummer, J. Harl, and G. Kresse,
Heyd-Scuseria-Ernzerhof hybrid functional for calculating the lattice dynamics of semiconductors,
\href{https://link.aps.org/doi/10.1103/PhysRevB.80.115205}{Phys. Rev. B \textbf{80}, 115205 (2009)}.
 
\bibitem{Hobbs2000}
D. Hobbs, G. Kresse, and J. Hafner,
Fully unconstrained noncollinear magnetism within the projector augmented-wave method,
 \href{https://link.aps.org/doi/10.1103/PhysRevB.62.11556}{Phys. Rev. B \textbf{62}, 11556 (2000)}.   

\bibitem{Kimchi2014}
I. Kimchi and A. Vishwanath,
Kitaev-Heisenberg models for iridates on the triangular, hyperkagome, kagome, fcc, and pyrochlore lattices,
 \href{https://link.aps.org/doi/10.1103/PhysRevB.89.014414}{Phys. Rev. B \textbf{89}, 014414 (2014)}. 

\bibitem{Catuneanu2015}
  A. Catuneanu, J. G. Rau, H. S. Kim, and H. Y. Kee, 
magnetic order proximal to the Kitaev limit in frustrated triangular systems: Application to ${\mathrm{Ba}}_{3}{\mathrm{IrTi}}_{2}{\mathrm{O}}_{9}$,
 \href{https://link.aps.org/doi/10.1103/PhysRevB.92.165108}{Phys. Rev. B \textbf{92}, 165108 (2015)}.

 
\bibitem{Kanamori1959}
J. Kanamori, 
Superexchange Interaction and Symmetry Properties of Electron Orbitals,
 \href{https://doi.org/10.1016/0022-3697(59)90061-7}{J. Phys. Chem. Solids \textbf{10}, 87-98 (1959)}. 

\bibitem{Goodenough1955}
J. B. Goodenough,
Theory of the Role of Covalence in the Perovskite-Type Manganites $[\mathrm{La}, M(\mathrm{II})]\mathrm{Mn}{\mathrm{O}}_{3}$,
 \href{https://link.aps.org/doi/10.1103/PhysRev.100.564}{Phys. Rev. B \textbf{100}, 564 (1955)}. 

\bibitem{SuppMat}
  See Supplemental Material at http://link.aps.org/supple
  -mental/10.1103/PhysRevB.000.000000
for the discussion of the HSE06 band compared to the PBE+$U$ band structures,
the magnetic coupling parameters of monolayer RuCl$_3$ calculated by spin spiral dispersion relations,
distribution of magnetic moments at different high symmetry points in the dispersion relation $E(\mathbf{q})$,
discussion of other potential spin models.
  

\bibitem{Riedl2022}
K. Riedl, D. Amoroso, S. Backes, A. Razpopov, T. P . T. Nguyen, K. Yamauchi, P. Barone, S. M. Winter, S. Picozzi and R. Valent\'{\i},
Microscopic origin of magnetism in monolayer $3d$ transition metal dihalides, 
\href{https://link.aps.org/doi/10.1103/PhysRevB.106.035156}{Phys. Rev. B \textbf{106}, 035156 (2022)}.
 
\bibitem{Ruderman1954}
M. A. Ruderman and C. Kittel,
Indirect exchange coupling of nuclear magnetic moments by conduction electrons,
 \href{https://link.aps.org/doi/10.1103/PhysRev.96.99}{Phys. Rev. \textbf{96}, 99 (1954)}.  

\bibitem{Kasuya1956}
T. Kasuya,
A theory of metallic ferro-and antiferromagnetism on Zener's model,
 \href{ https://doi.org/10.1143/PTP.16.45}{Prog. Theor. Phys. \textbf{16}, 45-57 (1956)}.  

\bibitem{Williams2015}
T. J. Williams, A. A. Aczel, M. D. Lumsden, S. E. Nagler, M. B. Stone, J.-Q. Yan, and D. Mandrus, 
Magnetic correlations in the quasi-two-dimensional semiconducting ferromagnet ${\text{CrSiTe}}_{3}$,
 \href{https://link.aps.org/doi/10.1103/PhysRevB.92.144404}{Phys. Rev. B \textbf{92}, 144404 (2015)}. 

\bibitem{Balcerzak2007}
T. Balcerzak,
A comparison of the RKKY interaction for the 2D and 3D systems and thin films,
 \href{https://doi.org/10.1016/j.jmmm.2006.10.494}{J. Magn. Magn. Mater. \textbf{310}, 1651-1653 (2007)}. 

\bibitem{Gong2017}
C. Gong, L. Li, Z. Li, H. Ji, A. Stern, Y. Xia, T. Cao, W. Bao, C. Wang, and Y. Wang
Discovery of intrinsic ferromagnetism in two-dimensional van der Waals crystals,
 \href{https://doi.org/10.1038/nature22060}{Nature \textbf{546}, 265-269 (2017)}. 

\bibitem{Marsman2002}
M. Marsman, and J. Hafner,
Broken symmetries in the crystalline and magnetic structures of \ensuremath{\gamma}-iron,
 \href{https://link.aps.org/doi/10.1103/PhysRevB.66.224409}{Phys. Rev. B \textbf{66}, 224409 (2002)}.  

\bibitem{Zhu2019}
Y. Zhu, Y. Pan, Z. Yang, X. Wei, J. Hu, Y. Feng, H. Zhang, and R. Wu,
Ruderman-Kittel-Kasuya-Yosida mechanism for magnetic ordering of sparse Fe adatoms on graphene,
 \href{https://doi.org/10.1021/acs.jpcc.8b11803}{J. Phys. Chem. C \textbf{123}, 4441-4445 (2019)}. 

\bibitem{Zhu2020}
Y. Zhu, Y. Pan, J. Fan, C. Ma, J. Hu, X. Wei, K. Zhang, and H. Zhang,
Strong phonon-magnon coupling of an O/Fe(001) surface,
 \href{https://doi.org/10.1007/s11433-020-1556-5}{Sci. China Phys. Mech. Astron. \textbf{63}, 117511 (2020)}.  

 \bibitem{Jiang2022}
L. Jiang, C. Huang, Y. Zhu, Y. Pan, J. Fan, K. Zhang, C. Ma, D. Shi, and H. Zhang, 
 Tuning the size of skyrmion by strain at the Co/Pt$_3$ interfaces,
 \href{https://doi.org/10.1016/j.isci.2022.104039}{Iscience \textbf{25}, 104039 (2022)}.

\bibitem{Huang2021}
C. Huang, L. Jiang, Y. Zhu, Y. Pan, J. Fan, C. Ma, J. Hu, and D. Shi, 
 Dzyaloshinskii-Moriya interaction via an electric field at the Co/h-BN interface,
 \href{https://doi.org/10.1039/D1CP02554F}{Phys. Chem. Chem. Phys. \textbf{23}, 22246-22250 (2021)}.  

\bibitem{Maksimov2020prr}
P. A. Maksimov and A. L. Chernyshev, 
Rethinking $\ensuremath{\alpha}\text{\ensuremath{-}}{\mathrm{RuCl}}_{3}$, 
 \href{https://link.aps.org/doi/10.1103/PhysRevResearch.2.033011}{Phys. Rev. Res. \textbf{2}, 033011 (2020)}.
 
\bibitem{Zhang2018}
X. Zhang, F. Mahmood, M. Daum, Z. Dun, J. A. M. Paddison, N. J. Laurita, T. Hong, H. Zhou, N. P. Armitage, and M. Mourigal,
Hierarchy of Exchange Interactions in the Triangular-Lattice Spin Liquid ${\mathrm{YbMgGaO}}_{4}$,
 \href{https://link.aps.org/doi/10.1103/PhysRevX.8.031001}{Phys. Rev. X \textbf{8}, 031001 (2018)}.

  \bibitem{Xu2020prb}
C. Xu, J. Feng, S. Prokhorenko, Y. Nahas, H. Xiang, and L. Bellaiche, 
 Topological spin texture in Janus monolayers of the chromium trihalides Cr(I, ${X)}_{3}$,
\href{https://link.aps.org/doi/10.1103/PhysRevB.101.060404}{Phys. Rev. B \textbf{101}, 060404(R) (2020)}.

 \bibitem{Stavropoulos2021prr}
  P. P. Stavropoulos, X. Liu, and H. Y. Kee, 
 Magnetic anisotropy in spin-3/2 with heavy ligand in honeycomb Mott insulators: Application to ${\mathrm{CrI}}_{3}$,
\href{https://link.aps.org/doi/10.1103/PhysRevResearch.3.013216}{Phys. Rev. Res. \textbf{3}, 013216 (2021)}.

\bibitem{Prange1979}
R. E. Prange and V. Korenman,
Local-band theory of itinerant ferromagnetism. IV. Equivalent Heisenberg model,
 \href{https://link.aps.org/doi/10.1103/PhysRevB.19.4691}{Phys. Rev. B \textbf{19}, 4691 (1979)}. 

\bibitem{Zhu2023}
Y. Zhu, Y. F. Pan, L. Ge, J. Y. Fan, D. N. Shi, C. L. Ma, J. Hu, and R. Q. Wu,
Separating RKKY interaction from other exchange mechanisms in two-dimensional magnetic materials,
\href{https://link.aps.org/doi/10.1103/PhysRevB.108.L041401} {Phys. Rev. B \textbf{108}, L041401 (2023)}.  
 


\end{thebibliography}

\newpage

\begin{thebibliography}{99}%

\bibitem{smXian2022}
J.-J. Xian, C. Wang, J.-H. Nie, R. Li, M. Han, J. Lin, W.-H. Zhang, Z.-Y. Liu, Z.-M. Zhang, M.-P. Miao, Y. Yi, S. Wu, X. Chen, J. Han, Z. Xia, W. Ji, and Y.-S. Fu,  Spin mapping of intralayer antiferromagnetism and field-induced spin reorientation in monolayer CrTe$_2$,
 \href{https://doi.org/10.1038/s41467-021-27834-z}{ Nat. Commun. \textbf{13}, 257 (2022)}.

\bibitem{smLi2021}
S. Li, S.-S. Wang, B. Tai, W. Wu, B. Xiang, X.-L. Sheng, and S. A. Yang, 
Tunable anomalous Hall transport in bulk and two-dimensional $1T\ensuremath{-}{\mathrm{CrTe}}_{2}$: A first-principles study,
 \href{https://link.aps.org/doi/10.1103/PhysRevB.103.045114}{Phys. Rev. B \textbf{103}, 045114 (2021)}.

 
\end{thebibliography}
\end{document}